\documentclass[pre,twocolumn,showpacs,nofootinbib,floatfix]{revtex4}
\usepackage{epsfig}
\usepackage{graphicx}
\usepackage{dcolumn}
\usepackage{bm}

\begin{document}
\title{How a ``Hit'' is Born: The Emergence of Popularity from the Dynamics of
Collective Choice}

\author{Sitabhra Sinha}%
\email{sitabhra@imsc.res.in}
\author{Raj Kumar Pan}%
\email{rajkp@imsc.res.in}

\affiliation{%
The Institute of Mathematical Sciences, C.I.T. Campus, Taramani,
Chennai - 600 113 India
}%
\date{\today}

\begin{abstract}
In recent times there has been a surge of interest in seeking out patterns
in the aggregate behavior of socio-economic systems. One such domain is the
emergence of statistical regularities in the evolution of collective choice
from individual behavior. This is manifested in the sudden emergence of
popularity or ``success'' of certain ideas or products, compared to their
numerous, often very similar, competitors.  In this paper, we present an
empirical study of a wide range of popularity distributions, spanning from
scientific paper citations to movie gross income. Our results show that in
the majority of cases, the distribution follows a log-normal form,
suggesting that multiplicative stochastic processes are the basis for
emergence of popular entities.  This suggests the existence of some general
principles of complex organization leading to the emergence of popularity.
We discuss the theoretical principles needed to explain this socio-economic
phenomenon, and present a model for collective behavior that exhibits
bimodality, which has been observed in certain empirical popularity
distributions.  
\end{abstract}

\maketitle

\section{Introduction}
\begin{small}
\begin{quote}
{\bf hit} ({\it noun}) a person or thing that is successful

{\bf popular} ({\it adj.}), from Latin {\it popularis}, from {\it populus}: 
the people, a people\\
{\bf 1:} of or relating to the general public,\\
{\bf 2:} suitable to the majority: as ({\bf a}) adapted to or indicative of the understanding and taste of the majority, ({\bf b}) suited to the means of the majority: inexpensive,\\
{\bf 3:} frequently encountered or widely accepted,\\
{\bf 4:} commonly liked or approved\\ 

\hfill {\em Merriam-Webster Online Dictionary}~\footnote{http://www.m-w.com/dictionary/}.
\end{quote}
\end{small}

In a pioneering study of how apparently rational people can behave
irrationally as part of a crowd, Charles MacKay~\cite{Mackay32} had given
several illustrations of certain phenomena becoming wildly popular without
discernible reason. In fact, he had focussed specifically on examples where
the individuals were behaving clearly contrary to their self-interest or
that of society as a whole, as for example, the habit of duelling or the
practise of witch-hunting. MacKay termed these episodes ``moral
epidemics'', long before the formal introduction of the concept of social
contagion~\cite{Morris00} and the use of biological epidemic models to
study such phenomena, ascribing their origin to the nature of men to
imitate the behavior of their neighbors.

However, such herding behavior~\footnote{MacKay referred to such behavior
as ``gregarious'', in its original sense of ``to flock''.} is not limited
to the examples given in MacKay's book, nor do the outcomes of such
behavior need to be so dramatic in their impact as, say, financial market
crashes or publicly sanctioned genocides. In fact, the sudden emergence of
a popular product or idea, that is otherwise indistinguishable in quality
from its competitors, is a more common example of the same process at work.
These events occur so often that we take such phenomena for granted;
however, the question of why certain products or ideas become much more
popular than what their intrinsic quality would warrant remains a
fascinating and unanswered problem in the social sciences.
Watts~\cite{Watts03} points this out when he says ``{\ldots} for every {\it
Harry Potter} and {\it Blair Witch Project} that explodes out of nowhere to
capture the public's attention, there are thousands of books, movies,
authors and actors who live their entire inconspicuous lives beneath the
featureless sea of noise that is modern popular culture.''

It may be worth mentioning that such popularity may be of different kinds,
one being runaway popularity immediately upon release, and, another being
modest initial popularity followed by ever-increasing popularity in
subsequent periods.  The former is thought to be driven by the advertising
blitz preceding the release or launch of the product while the latter has
sometimes been explained in terms of self-reinforcing effects, where a
slight relative edge in terms of initial popularity results in more
consumers being inclined towards the slightly more popular product, thereby
increasing its popularity even further and so on, driving up its popularity
through positive feedback.

As physicists we are naturally interested to see whether there are general
trends that can be observed in popularity phenomena across a large range of
contexts in which they are observed. An allied question is whether this
popularity can be related to any of the intrinsic properties of the
products or ideas, or whether this is entirely an outcome of a sequence of
chance events. The fact that often popular products are seen to be not all
that qualitatively different from their competitors, or in some cases,
actually somewhat inferior, seems to weigh against the former possibility.
However, we would like to see whether the empirically observed popularity
distributions also suggests the latter alternative. We also need to see
whether pre-release advertising does indeed play a role in creating a high
initial burst of popularity.

In this article, we first approach the problem empirically, looking at
previous work done on measuring popularity distributions, as well as
presenting some of our recent analysis of the popularity phenomena
occurring in a variety of different contexts.  One remarkable universality
we find is that most popularity distributions we examine seem to have long
tails, and can be fit either by a log-normal or a power-law probability
distribution function, the exponent of the latter often being quite close
to $-2$. Another interesting feature observed for some distributions is
their bimodal character, with the majority of instances occurring at
extreme ends of the distribution, while the center of the distribution is
remarkably under-represented. Both of these features indicate a significant
departure from the Gaussian distribution that may have been naively
expected.  Next, we survey possible theoretical models for explaining the
above features of the empirical distributions. In particular, we discuss
how log-normal distributions can arise through several agents making
independent decisions in choosing from a range of products with randomly
distributed qualities. We also present a model of agent-agent interaction
that shows a transition from unimodal to bimodal distribution of the
collective choice, when agents are allowed to learn from their previous
experience.  We conclude with a short discussion on how log-normal and
power-law tail distributions can be generated from the same theoretical
framework, the former occurring when agents choose independently of other
agents (basing their decisions on individual perceptions of quality) and
the latter emerging when agent-agent interactions are crucial in deciding
the desirability of a product.

\section{Empirical Popularity Distributions}
In studying the popularity distribution of products, the first question one
needs to resolve is how to measure popularity. While in some cases this may
seem rather obvious, e.g., the number of people buying a particular book,
in other cases it may be difficult to identify a unique measure that will
satisfy everyone. For example, the popularity of movies can be measured
either in terms of an average over critics' opinions published in major
periodicals, web-based voting in movie-related online communities, the
income generated when a movie is running in theaters, or the cumulative
sales and rentals from DVD stores. In most cases, we have let the quality
of the available data decide our choice of which popularity measure to use.

An equally important question one needs to answer is the nature of the
statistical distribution with which to fit the data. In almost all cases
reported below, we observe distributions that deviate significantly from
the Gaussian distribution in having extremely long tails. The occurrence of
such fat-tailed distributions in so many instances is very exciting, as it
indicates that the process of emergence of popular products is more than
just $N$ agents independently making {\em single} binary (i.e., {\em yes}
or {\em no}) decisions to adopt a particular choice. However, to go beyond
this conclusion and to identify the possible process involved, one needs to
ascertain accurately the true nature of the distribution.  This brings up
the question of how to obtain the probability density function (PDF) from
the empirical data. The method generally used is to arrange the data into a
suitable number of bins to obtain a histogram, which in an appropriate
limit will provide the PDF. This works fine when the underlying
distribution is Gaussian with sharply decaying tails; however, for
long-tailed distributions, it is exactly the extreme ends one is interested
in, which have the least representation in the data. As a result, the PDF
is extremely noisy at the tails, and hence, it is often hard to conclude
the nature of the distribution. Often, one can remove some of the noise by
using the PDF to generate the cumulative distribution function (CDF), which
is essentially the probability that an event is larger than a given
size~\footnote{The CDF, $P_c ( x )$, of a given process is obtained by
integrating the corresponding PDF, $P ( x )$, i.e., $P_c ( x ) =
\int_{x}^{\infty} P (x^{\prime}) dx^{\prime}$.}.  As larger quantities of
data points are now accumulated in each of the bins, the tail becomes
smoother in the CDF plot.  However, the data binning process is susceptible
to noise, that can change significantly the shape of the distribution,
depending on the size and boundary values of each bin.  This can lead to
serious errors, e.g., wrongly identifying the tail of the distribution to
be following a power law. Even if the distribution indeed has a power-law
tail, one may obtain a quantitatively erroneous value for the power-law
exponent by using graphical methods based on linear least square fitting on
a double logarithmic scale~\cite{Goldstein04}. 

A better way to examine the nature of the tail of a distribution is to
avoid binning altogether and to switch to a rank-ordered plot of the data,
which allows one to focus on the upper tail of the distribution containing
the data points of largest magnitude. These plots are often referred to as
{\em Zipf plots}, after the Harvard linguist, G.~K.~Zipf, who used such
rank-frequency plots of the occurrence of the most common words in the
English language to establish a scaling relation for written natural
languages~\cite{Zipf32,Zipf49}.  In this procedure, the data points are
ranked or arranged in decreasing order of their magnitude.  Note that the
CDF can be obtained from the rank-ordered plot by simply exchanging the
abscissae and the ordinate, and suitably scaling the axes. Thus, by
avoiding binning one can make a better judgement of the nature of the
distribution.  To quantitatively determine the parameters of the
distribution, one of the most robust methods is maximum likelihood
estimation (MLE)~\cite{Newman05}.  For example, if the underlying
distribution $P_c ( x )$ has a power-law tail, then the CDF exponent can be
obtained from the MLE method by using the formula 
\begin{equation}
\alpha=n\sum_{i=1}^{n}\left[{\rm ln} {\frac{x_i}{x_{min}}}\right]^{-1},
\label{ss:MLE}
\end{equation}
where, $x_{min}$ corresponds to the minimum value of $x$ for which the
power-law behaviour holds. Similarly, one can obtain maximum likelihood
estimates of the parameters for log-normal and other distributions. 

It is, of course, obvious that the results from the three different plots,
namely, the PDF, the CDF and the rank-ordered, should be related to each
other. So, for example, if the CDF of an empirically obtained distribution
is found to exhibit a power-law tail which can be expressed as, 
\begin{equation}
P_c (x) \sim x^{- \alpha},
\label{ss:CDF}
\end{equation}
with the characteristic exponent~\footnote{This exponent $\alpha$ is often
referred to as the {\em Pareto exponent}, after the Italian economist,
V.~Pareto, who was the first to report power law tails for the CDF of
income distribution across several European countries~\cite{Pareto97}.}
$\alpha$, it is easy to show that the PDF and the rank-ordered plots will
also exhibit power-law behavior~\cite{Adamic02}.  Moreover, the exponents
of the power-law seen in these two cases will be related to the
characteristic exponent of the CDF, $\alpha$, as follows: the PDF will
follow the relation, 
\begin{equation}
P(x) \sim x^{- (\alpha + 1)},
\label{ss:PDF}
\end{equation}
while, the rank-ordered plot will exhibit the relation,
\begin{equation}
x_k \sim k^{- 1/\alpha},
\label{ss:rankorder}
\end{equation}
where $x_k$ denotes the $k$-th ranked data point. The above examples are
all given for the case when the underlying distribution has a power-law
tail; similar relations can be derived for other underlying distributions,
e.g., log-normal. 
 
\subsection{Examples}
In the following paragraphs we have briefly surveyed previous empirical
work on popularity distribution, as well as, presented some of our own
recent analysis of popularity data from a broad variety of contexts. In
most cases, we have characterized the empirical CDF with a log-normal fit
over the entire distribution.  However, in those cases where the data is
available only for the upper tail of the distribution, such a procedure is
not possible.  In these cases, we have presented a rank-ordered plot of the
data and have tried to fit a power-law characterized by the CDF exponent,
$\alpha$.  In this context, we note that most previous observations of
popularity distributions had focussed on the upper tail, and fitted a
power-law on this. However, we find that the entire distribution is very
often a much better fit to the log-normal distribution~\cite{Limpert01}.
We conclude with a brief discussion of why data that fit log-normal much
better has often been reported in the literature to follow a power-law
tail. 

\subsubsection{City Size.}
Possibly the first ever empirical observation of a long-tailed popularity
distribution is that of cities, as measured by their population, which was
first proposed in 1913 by Auerbach~\cite{Auerbach13}.  Later, this basic
idea was refined by many others, most notably Zipf~\cite{Zipf49}. In fact,
the last mentioned work has become so well-known that, often the term {\em
Zipf's law} is used to refer to the idea that city sizes follow a
cumulative probability distribution having a power-law tail~\cite{Li03}
with exponent $\alpha = 1$. Over the years, several empirical studies have
been published in support of the validity of Zipf's law~\cite{Gabaix99}.
However, other empirical studies have found significant deviations from the
exact form given by Zipf~\cite{Soo05}.  In a recent review, the combined
estimate of the exponent $\alpha$ from 29 different studies is found to be
significantly larger than 1 suggesting a less extended tail than implied by
a strict interpretation of Zipf's law~\cite{Nitsch05}.  All these studies
have focused on the upper tail ({\em i.e.}, larger cities) of the
distribution. If one also considers the smaller cities, the whole
distribution often fits a double-Pareto log-normal, i.e., a distribution
which is log-normal in the bulk but has long tails at the two
ends~\cite{Reed02}.  Even the power-law fit of the tail has itself been
called into question by a study of the size distribution of US cities over
the period 1900-1990~\cite{Black03}. These results are of special
significance to our study, as it shows that the fat-tailed distribution of
popularity of cities need not be a power-law but could be explained by
other distributions.

\subsubsection{Company Size.}
Almost of similar vintage to the city size literature is the work on
company size, measured in terms of sales or employees. Note that, both of
these are measures of popularity of the company, the former measuring its
popularity among the consumers of its products, while the latter measures
its popularity in the labor market.  In 1932, Gibrat formulated the {\em
law of proportional growth}, essentially a multiplicative stochastic
process for explaining company growth, which predicts that the distribution
of firm size would follow a log-normal
distribution~\cite{Gibrat32,Sutton97}. While this has indeed been reported
from empirical data~\cite{Stanley95,Cabral03}, there have been also reports
of a power-law tail~\cite{Ramsden00}. In particular, Axtell~\cite{Axtell01}
has looked at the size of US companies (listed in the U.S. Census Bureau
database) in terms of the number of employees, that yields a CDF with power
law tail whose exponent $\alpha \sim 1$. When the size was expressed in
terms of receipts (in dollars) this also yielded a power law CDF with
$\alpha \sim 0.99$. 

\subsubsection{Scientists and Scientific Papers.}
The study of popularity in the field of science has a rich and colorful
history~\cite{Hagstrom65}.  One of the earliest such studies is that on the
visibility of scientists, as measured by subjective opinions elicited from
a sample of the scientific community~\cite{Cole68}.  The skewed nature of
the visibility because of misallocation of credit in the field of science,
where an already famous scientist gets more credit than is due compared to
less well-known colleagues, has been termed as the {\em Mathew
effect}~\cite{Merton68}. This is quite similar to the unequal degree of
popularity seen in show-business professions, e.g., among movie actors and
singers. A more objective measure for the popularity of scientists is the
total number of citations to their papers~\cite{Sornette98}. 

\begin{figure}
\includegraphics[width=.85\linewidth]{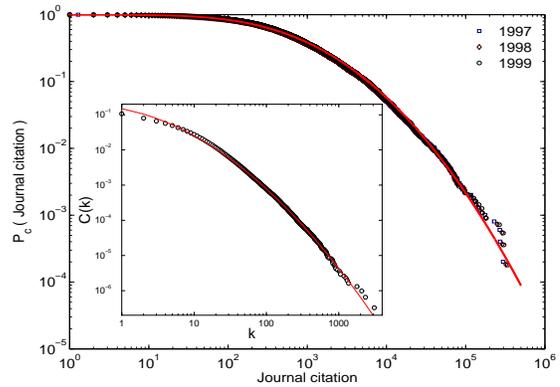}
\caption{The cumulative distribution function for the total number of 
citations to a journal in a given year, for all journals ($\sim 5500$) 
listed in ISI Journal Citation Report (Science edition) 
for the years 1997-1999, fit by a log-normal curve ({\em in red}).
The inset (from Ref.~\cite{Redner05}) 
shows the cumulative probability distribution of citations, 
$C(k)$, against the number of citations, $k$, to all papers published from 
July 1893 through June 2003 in the Physical Review journals, fit by a 
log-normal curve ({\em in red}).}
\label{ss:impact}
\end{figure}

The popularity of individual scientific papers can also be analysed in
terms of citations to them~\cite{Price65}. Price~\cite{Price76} had tried
to give a theoretical model based on {\em cumulative advantage} along with
supporting evidence showing that the distribution of citations to papers
follow a power-law tail.  More recently, in a study~\cite{Redner98}
analyzing papers in the Institute for Scientific Information (ISI)
database, as well as papers published in {\em Physical Review D}, Redner
concluded that the probability distribution of citations follow a power law
tail with an exponent close to $-3$.  However, in a later work looking at
all papers published in {\em Physical Review} journals over the past 110
years, this distribution was found to be fit better by a
log-normal~\cite{Redner05} (Fig.~\ref{ss:impact}, inset).  

In addition to the popularity of individual papers measured by the number
of their citations, one can also define the popularity of the journals in
which these papers are published by considering the total number of
citations to all articles published in a journal.  In Fig.~\ref{ss:impact},
we have plotted the cumulative distribution of the total citations in
1997-99 to all papers ever published in a journal. The data has been fit
with a log-normal distribution; maximum likelihood estimates of parameters
for the corresponding distribution are $\mu=6.37$ and $\sigma=1.75$.

\subsubsection{Newspaper and Magazines.}
\begin{figure}
\includegraphics[width=.85\linewidth]{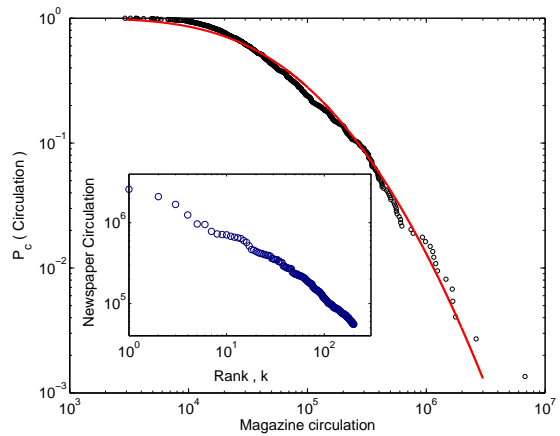}
\caption{
Cumulative distribution function of the 740 most circulated
magazines in UK, fit by a log-normal curve ({\em in red}).  
The inset shows the rank-ordered plot of the top 200 newspapers in USA
according to circulation.}
\label{ss:newsmag}
\end{figure}

The popularity of scientific journals naturally leads us to wonder about
the popularity distribution for general interest magazines as well as
newspapers. An obvious measure of popularity in this case is the
circulation figure. Fig.~\ref{ss:newsmag} shows the CDF of the top 740
magazines according to average net circulation per issue in the United
Kingdom~\footnote{http://www.abc.org.uk} in 2005.  The figure shows an
approximately log-normal fit; maximum likelihood estimates of parameters
for the corresponding distribution are $\mu=10.79$ and $\sigma=1.18$.
Next, we analyzed the circulation figures for the top 200 newspapers in the
USA for the year 2005 according to their
circulation~\footnote{http://www.accessabc.com/reader/top150.htm}.
Fig.~\ref{ss:newsmag}(inset) shows the corresponding rank-ordered plot with
an approximate power-law fit over a decade yielding Zipf's law, which is
supported by the maximum likelihood estimate of the exponent for the
cumulative probability density function, $\alpha \sim 1.12$.  

\subsubsection{Movies.}
\begin{figure}
\includegraphics[width=.85\linewidth]{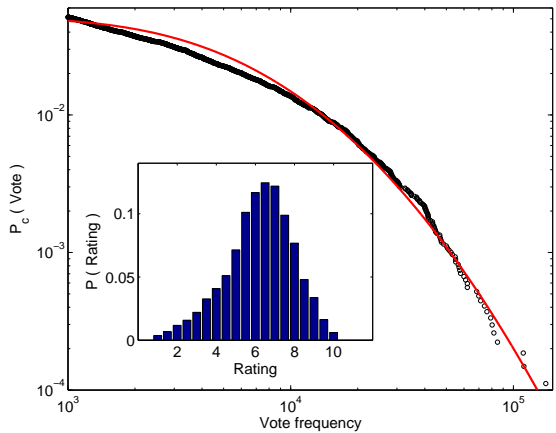}
\caption{Cumulative probability distribution of the number of votes given
by registered users of IMDb to movies and TV series released or shown
between the years 2000-2004, fit by a log-normal curve ({\em in red}).
(Inset) The probability distribution of the IMDb rating of a movie,
averaged over all the votes received.}
\label{ss:movievotes}
\end{figure}

Movie popularity can be measured in a variety of ways, e.g., by looking at
the votes given by users of various movie-related online forums. One of the
largest of such forums is the Internet Movie Database
(IMDb)~\footnote{http://www.imdb.com} that allows registered users to rate
films (and television shows) in the range 1-10 (with 1 corresponding to
``awful'' and 10 as ``excellent'').  We looked at the cumulative
distribution of all votes received by movies or TV series shown between
2000-2004 (Fig.~\ref{ss:movievotes}).  The tail of the distribution
approximately fits a log-normal distribution, with maximum likelihood
estimates of the corresponding parameters, $\mu=8.60$ and $\sigma=1.09$.
Next, we look at the distribution of average rating given to these items.
As the minimum and maximum ratings that an item can receive are 1 and 10,
respectively, this distribution is necessarily bounded. The skewed
probability distribution of the average rating resulting from our analysis
is shown in Fig.~\ref{ss:movievotes}~(inset). 
\begin{figure}
\includegraphics[width=.85\linewidth]{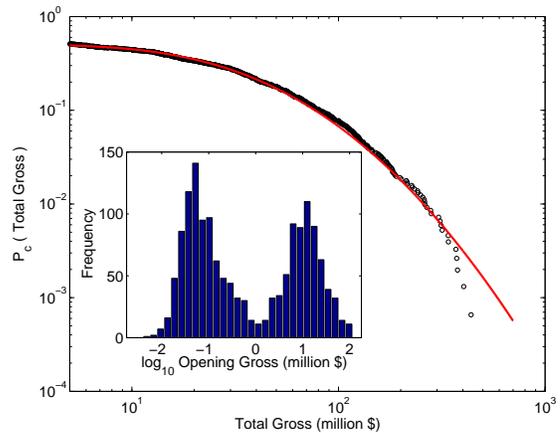}
\caption{Cumulative distribution of total gross income for movies released 
across theaters in USA during 2000-2004, fit by a log-normal curve 
({\em in red}). The inset shows the distribution of movie income 
according to the opening weekend gross.}
\label{ss:movieincome}
\end{figure}

\begin{figure*}
\includegraphics[width=.4\linewidth]{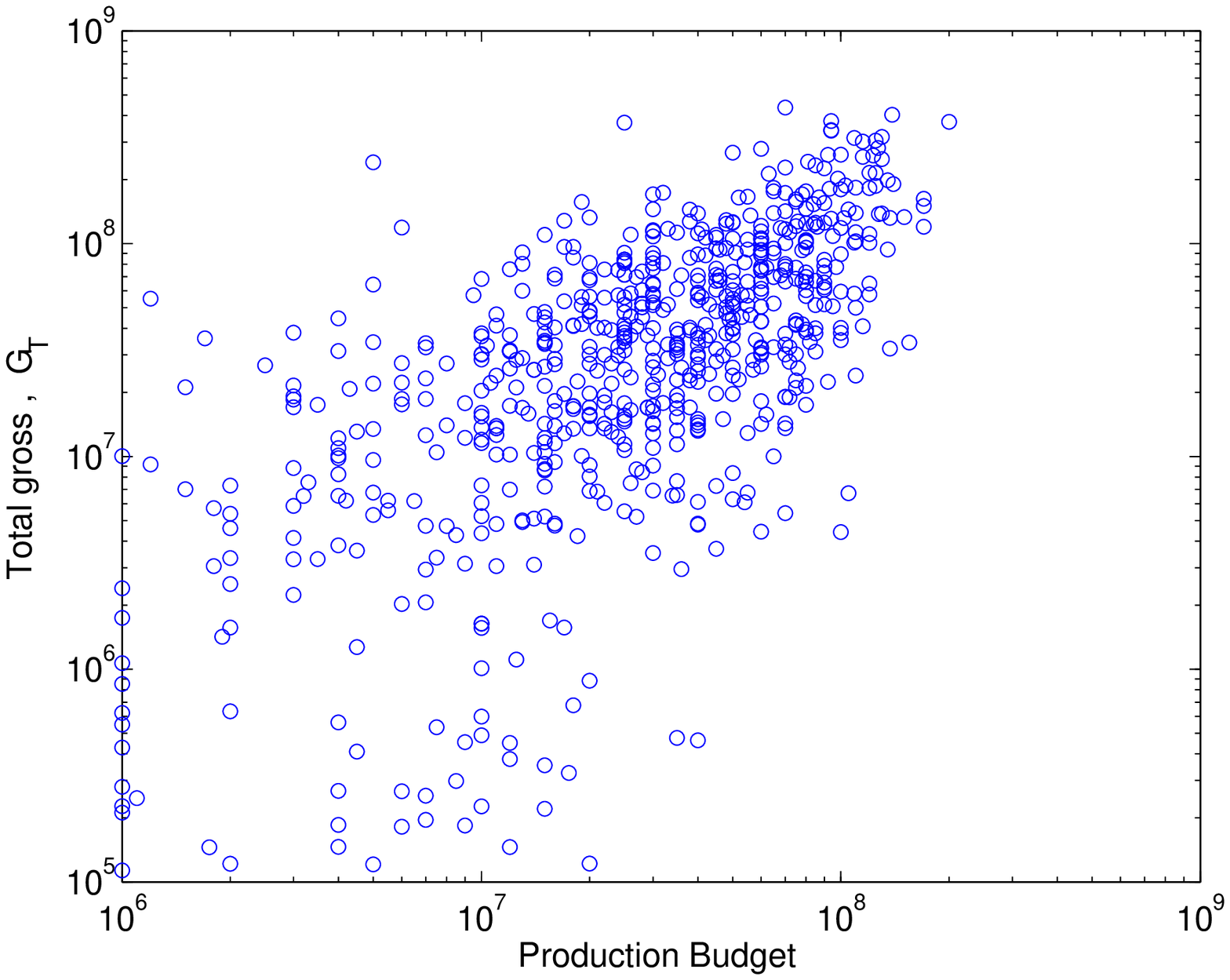}
\includegraphics[width=.4\linewidth]{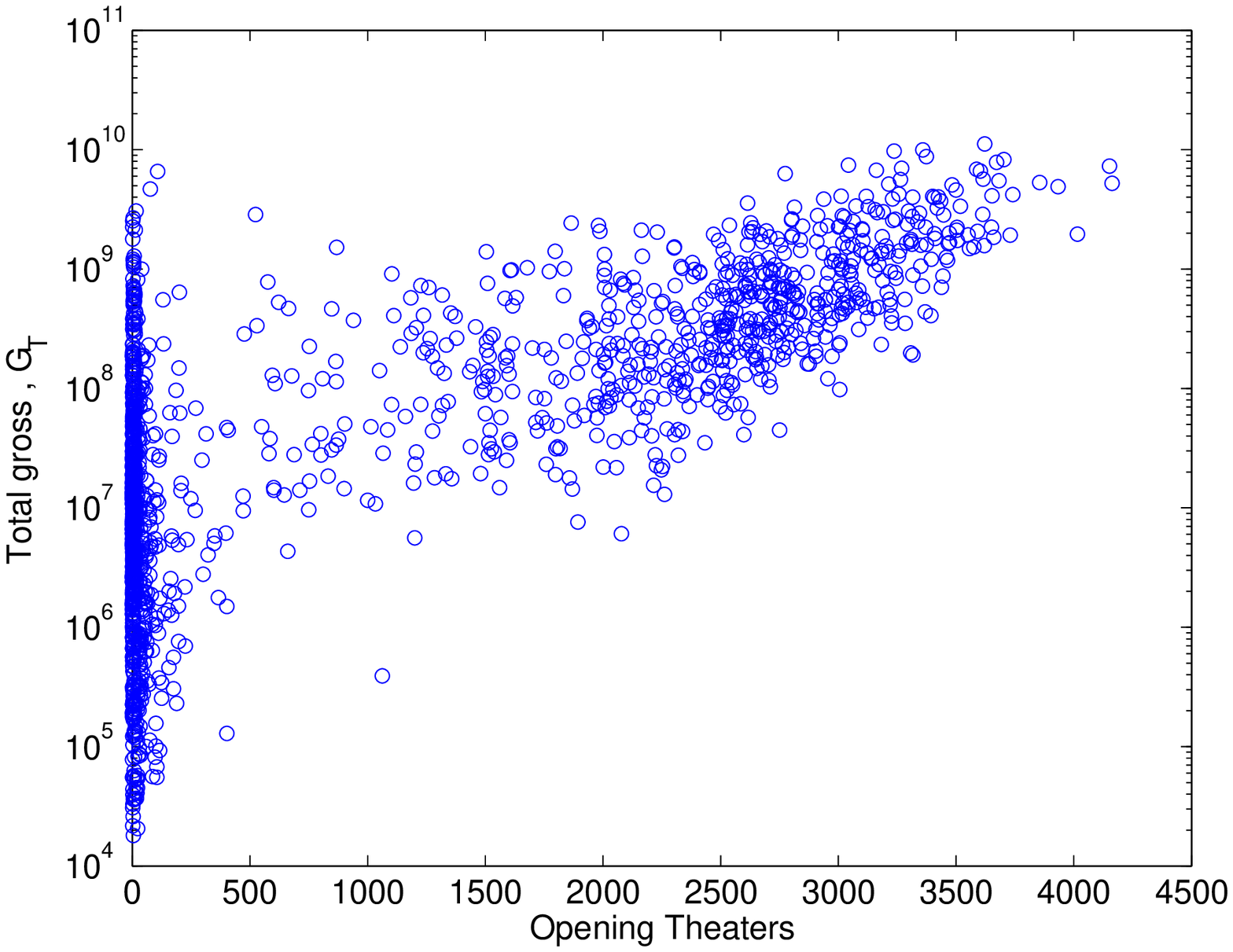}
\caption{(Left) The total gross ($G_T$, in dollars) of a movie vs its
production budget (in dollars).  (Right) The total gross ($G_T$, in
dollars) of a movie vs the number of theaters it is released on the opening
weekend.}
\label{ss:budgetadvt}
\end{figure*}
The measures used above have many drawbacks as indicators of movie
popularity, particularly so when they are aggregated to produce average
values. For example, users may judge different movies according to very
different information, with so-called classic movies faring very
differently from recently released movies that have very little information
available about them. Also, it does not cost anything to vote for a movie,
so that the vital element of competition among movies to become popular is
missing in this measure. In contrast, looking at the gross income
distribution of movies that are being shown at theaters gives a sense of
the relative popularity of movies that have roughly equal amount of
information available about them. Also, this kind of ``voting with one's
wallet'' is a truer indicator of the viewer's movie preferences.  The
freely available datasets about weekly earnings of most movies released
across theaters in the USA makes this a practical exercise.  For our study
we have concentrated on data from {\em The Movie
Times}~\footnote{http://www.the-movie-times.com} and {\em The
Numbers}~\footnote{http://www.the-numbers.com/} websites for the period
2000-2004.  Although total gross may be a better measure of movie
popularity, the opening gross is often thought to signal the success of a
particular movie. This is supported by the observation that about 65-70
$\%$ of all movies earn their maximum box-office revenue in the first week
of release~\cite{Vany99}.  The rank-ordered distribution for the opening,
as well as the total gross, show an approximate power law with an exponent
$1/ \alpha \sim - 1/2$ in the region where the top grossing movies are
located~\cite{Sinha04a}.  However, when the data are aggregated together we
find that the distribution (Fig.~\ref{ss:movieincome}) is better fit by a
log-normal~\footnote{We have also verified this for the income distribution
of Indian movies.} (similar to the observation of Redner vis-a-vis
citations)~\cite{Pan06}. The maximum likelihood estimates of the log-normal
distribution parameters yield $\mu=3.49$ and $\sigma=1.00$.  Further, we
observe that the total gross distribution is just a scaled version of the
opening distribution, which essentially implies that the popularity
distribution of movies is decided at the opening itself.  An additional
feature of interest is that both the opening and the total gross
distributions are bimodal (Fig.~\ref{ss:movieincome}, inset), implying that
most movies either do very well or very badly at the box office.

We have tried to see whether the popularity of individual movies correlate
with its production quality (as measured by production budget).
Fig.~\ref{ss:budgetadvt}~(left) shows a plot of the total gross vs
production budget for a large number of movies released between 2000-04
whose budget exceeded $10^6$ \$. As is clear from the figure, although in
general, movies with higher production budget tend to earn more, there is
no significant correlation (the correlation coefficient is only 0.62).  One
can also argue that the determination of success of a movie on its opening
implies the key role of pre-release advertising. Although the data for
advertising budget is often unavailable, we can use as a surrogate, the
data about the number of theaters that a movie is initially released at,
since the advertising cost will scale with this quantity. As is obvious
from Fig.~\ref{ss:budgetadvt}~(right), the correlation here is worse,
indicating that advertising has often very little role to play in deciding
the success or otherwise of a movie in becoming popular. In this context,
one may note that De Vany \& Walls have looked at the distribution of movie
earnings and profit as a function of a variety of variables, such as,
genre, ratings, presence of stars, etc. and have not found any of these to
be significant determinants~\cite{Vany03}. 

To make a quantitative analysis of the relative performance of movies, we
have defined the persistence time $\tau$ of a movie as the time (measured
in number of weekends) upto which it is being shown at theaters.  We
observe that most movies run for upto about 10 weekends, after which there
is a steep drop in their survival probability. The empirical data seem to
fit a Weibull distribution quite well. 

\subsubsection{Websites and Blogs}
Zipf's law for the distribution of requests for pages from the web was
first reported by Glassman~\cite{Glassman94}. By tracing web accesses from
DEC's Palo Alto facilities, $~ 10^5$ HTTP requests were gathered and the
rank-ordered distribution of pages was shown to have an exponent $\sim -1$.
This was supported by a popular article~\cite{Nielsen97} which observed
Zipf's law when analysing the incoming page-requests to a single site
(www.sun.com).  However, subsequent investigation of the page request
distribution seen by web proxy caches using traces from a variety of
sources, found the  rank-order exponent  to vary between 0.64 to
0.83~\cite{Breslau99}.  The deviation from the earlier result (showing
exact Zipf's law) was ascribed to the fact that web accesses at a web
server and those at a web proxy are different, because the former includes
requests from all users on the Internet while the latter includes only
those users from a fixed group.  Access statistics for web pages have also
been analysed by Adamic and Huberman from the access logs of about 60000
individual usage logs from America Online~\cite{Adamic00}. The resulting
cumulative distribution of website popularity, according to the number of
unique visits to a website by users, showed a power law fit with $\alpha$
very close to 1. 

Another obvious measure of webpage popularity is the number of links to it
from another webpage. Distribution of incoming links to a webpage (i.e.,
URLs pointing to a certain HTML document) for the nd.edu domain, have been
shown to obey a power law with exponent $\simeq -2.1$~\cite{Albert99}. This
power law was quantitatively confirmed (i.e., the same exponent value of
2.1 was reported) over a much larger data set involving a web-crawl on the
entire WWW with $2 \times 10^8$ webpages and $1.5 \times 10^9$
links~\cite{Broder00}. While the power law distribution of popularity of
websites according to the number of incoming links has been
well-established as a power law, among web-pages of the same type (e.g.,
the set of US newspaper homepages) the bulk of the distribution of incoming
links deviates strongly from a power law, exhibiting a roughly log-normal
shape~\cite{Pennock02}.

The finding that the micro-structure of popularity within a group is closer
to a log-normal distribution has created some controversy among researchers
involved in measuring the popularity distribution of blogs~\footnote{A blog
or weblog has been defined as a web page with minimal to no external
editing, providing on-line commentary, periodically updated and presented
in reverse chronological order, with hyperlinks to other online
sources~\cite{Drezner04}. Blogs can function as personal diaries, technical
advice columns, sports chat, celebrity gossip, political commentary, or all
of the above.} which have over the past few years picked up a large
following all over the web. Shirky~\cite{Shirky03} had arranged 433
weblogs in rank order according to number of incoming links from other
blogs and had claimed an approximate power law distribution. In contrast
to this, Drezner \& Farrell~\cite{Drezner04} conducted a study of the
incoming link distribution of over 4000 blogs dealing almost exclusively
with political topics, and found the distribution to be much better fit by
a log-normal than a power law.  Other studies have made contradictory
claims about whether the popularity of blogs is better fit by a log-normal
or power-law tailed distribution~\cite{Hindman03,Adamic05}.

\begin{figure}
\includegraphics[width=.85\linewidth]{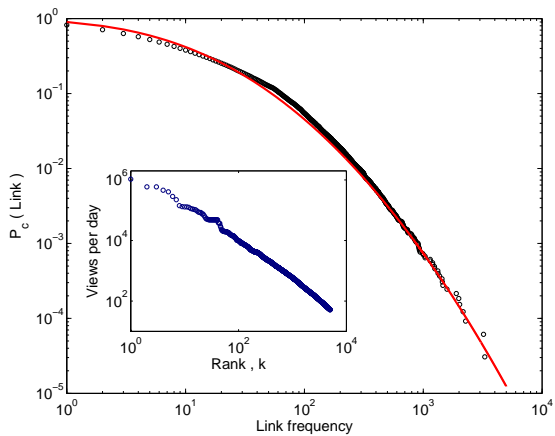}
\caption{Cumulative distribution function for blog popularity measured by
the number of incoming links a blog receives from other weblogs listed in
the TTLB Blogosphere ecosystem within the past 7-10 days.  The curve is the
best log-normal fit to the data.  The inset shows the rank-ordered plot of
blog popularity according to the number of visits to a blog in a single
day, in the TTLB ecosystem.}
\label{ss:blogs}
\end{figure}
We have also analysed the popularity distribution of blogs according to
citations in other blogs, using three different blogosphere ecologies,
i.e., directories of blog listings. Such ecologies scan all blogs
registered with them for (i) the number of links they receive from other
blogs in their list, as well as (ii) the number of visits to that blog.
These two measures of popularity complement each other, as the former looks
at who is getting the most links from other bloggers, while the latter
shows which blogs are actually receiving the most readers.  The most
extensive data that we have analyzed comes from the TTLB Blogosphere
ecosystem~\footnote{http://truthlaidbear.com/} that lists 52048 blogs.  In
Fig.~\ref{ss:blogs} we show the CDF for the popularity of blogs from this
ecology, measured from the number of links to that blog seen in the ``front
page'' of other member blogs within the past 7-10 days. This can be
considered a rolling snapshot of the relative popularity of different blogs
at a particular instant of time.  For comparison, we also looked at data
from two other ecologies, namely, the
Technorati~\footnote{http://www.technorati.com/} and the
Blogstreet~\footnote{http://www.blogstreet.com/} ecosystems, and observed
qualitatively almost identical behavior.  The CDF (Fig.~\ref{ss:blogs})
shows an approximately log-normal fit; maximum likelihood estimates of
parameters for the corresponding distribution are $\mu=1.98$ and
$\sigma=1.51$.  We have also analyzed the popularity of blogs listed in the
TTLB ecosystem according to traffic, i.e., views per day
(Fig.~\ref{ss:blogs}, inset), which shows a power law over almost two
decades for the rank-ordered plot. The maximum likelihood estimate of the
corresponding exponent for the cumulative probability density yields
$\alpha \sim 0.67$. 

\subsubsection{File Downloads.}
\begin{figure*}
\includegraphics[width=.4\linewidth]{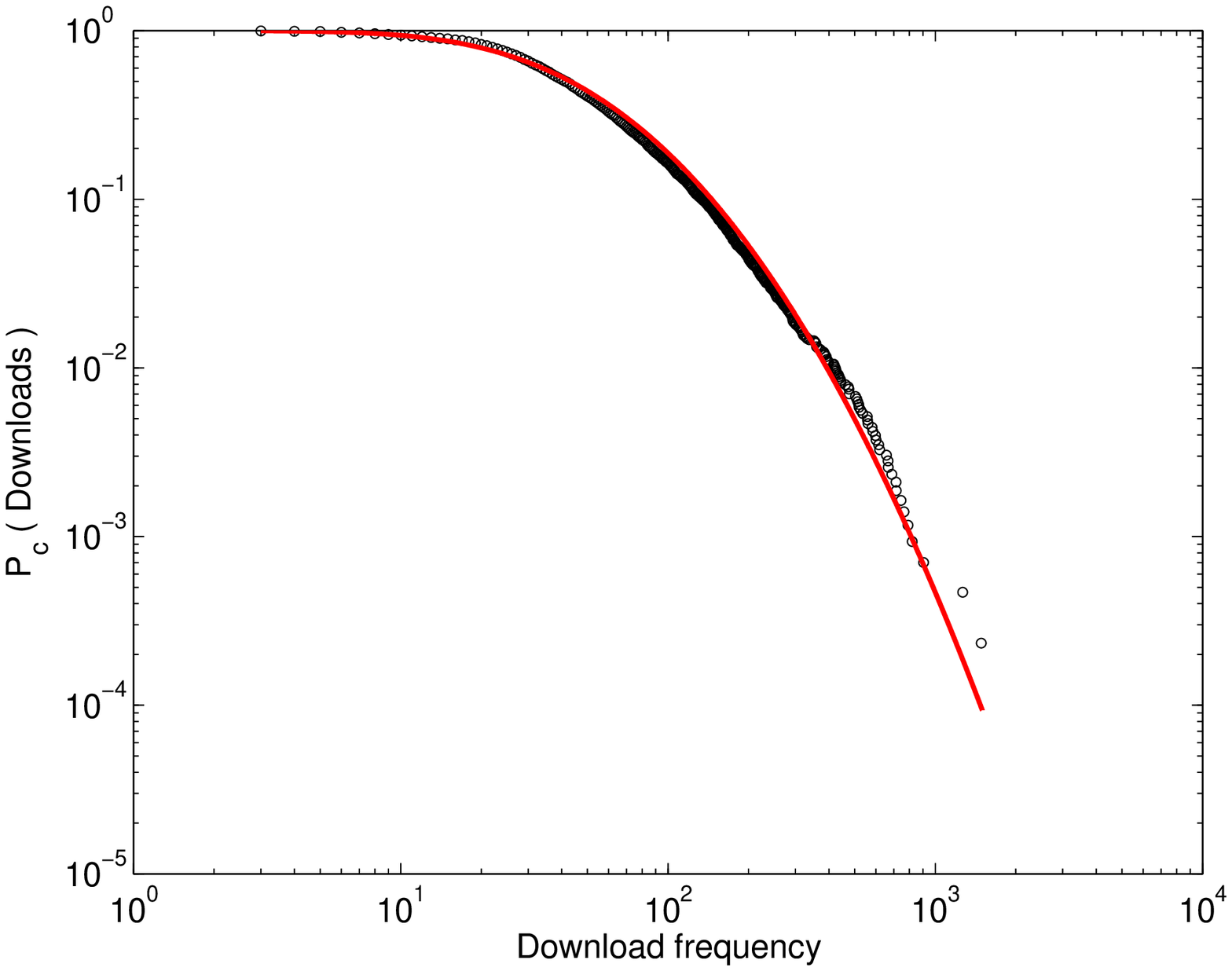}
\includegraphics[width=.4\linewidth]{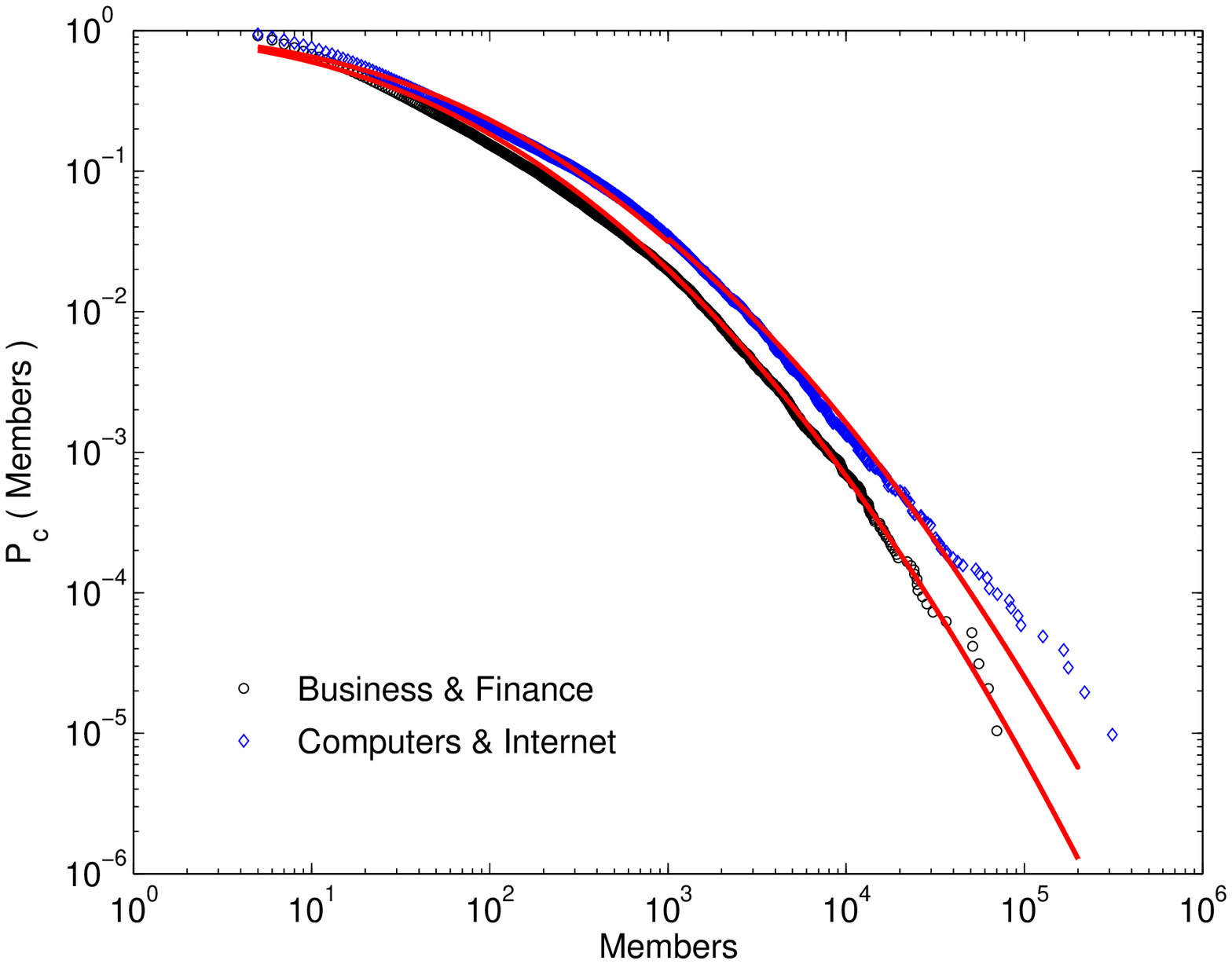}
\caption{(Left) Cumulative distribution function for the number of
downloads of different files in 1 month during early 2006 from the MATLAB
file exchange site.  (Right) Cumulative distribution function for the
number of members in different Yahoo groups under the Business \& Finance
(squares) and Computers \& Internet (diamonds) categories.  Groups with
less then 5 members are not considered.  For both figures, the curves ({\em
in red}) are the best log-normal fits to the data.}
\label{ss:filesgroups}
\end{figure*}

Another web-related measure of popularity is that of file downloads. There
are numerous file repositories in the net which allow visitors to download
files either freely or for a fee. We focussed on files stored in the MATLAB
Central File
Exchange~\footnote{http://www.mathworks.com/matlabcentral/fileexchange/},
which are computer programs.  We looked at the number of downloads of all
files over a period of one month during early 2006. The CDF
[Fig.~\ref{ss:filesgroups}~(left)] shows an approximately log-normal fit;
maximum likelihood estimates of parameters for the corresponding
distribution are $\mu=3.76$ and $\sigma=0.89$. 

\subsubsection{Groups.}
A fertile area for observing the distribution of popularity is in the arena
of social groups. While the membership of clubs, gangs, co-operatives,
secret societies, etc., are difficult to come by, with the rising
popularity of the internet it is easy to obtain data for online communities
such as those in Yahoo~\footnote{http://groups.yahoo.com} or
Orkut~\footnote{http://www.orkut.com}. By observing the memberships of
each of the groups in the community that a user can join, one can have a
quantitative measure of the popularity of these groups. An analysis of the
Yahoo groups resulted in a fat-tailed cumulative distribution of the group
size~\cite{Noh05}. Even though the distribution has a significant curvature
over the entire range, the tail fits a power law for slightly more than a
decade, with exponent $\alpha = 1.8$.

We have recently carried out a smaller-scale study of the popularity of
Yahoo groups~\footnote{The entire Yahoo groups community is divided into 16
categories, each of which are then further divided into subcategories.}. As
in the earlier study, the popularity of the groups in each category has
been estimated by the number of group members.
Fig.~\ref{ss:filesgroups}~(right) looks at the cumulative distributions of
the group size for two categories, namely Business \& Finance and Computer
\& Internet, which comprise 182086 and 172731 groups respectively.
However, unlike the power-law reported in the earlier study, we found both
the distributions to approximately fit a log-normal form, with the
parameters for the corresponding distributions being $\mu=2.80$,
$\sigma=2.00$ and $\mu=3.10$, $\sigma=2.05$, respectively.

One can also look at the popularity of individual members of an online
group, which has been analysed for a different type of community in the
web: that formed by the users of the {\it Pretty-Good-Privacy} (PGP)
encryption algorithm. To ensure that identities are not forged, users
certify one another by ``signing'' the other person's public encryption
key. In this manner, a directed network (the ``web of trust'') is created
where the vertices are users and links are the user certifications. A
measure of popularity in this case will be the number of certifications
received by an user from other users, i.e., the number of incoming links
for a vertex in the ``web of trust''. The in-degree cumulative distribution
has been reported to be a power law with the exponent $\alpha \simeq
1.8$~\cite{Guardiola02}. 

\subsubsection{Elections.}
Political elections are processes that can be viewed as contests of
popularity between individual candidates, as well as parties. The fraction
of votes received by candidates is a direct measure of their popularity,
regardless of whether the electoral system uses a majority voting rule
(where the candidate with the largest number of votes wins) or a
proportional representation (parties getting representation at the
legislative house proportional to their fraction of the popular vote). Such
studies have been carried out for, e.g., the 1998 Brazilian general
elections~\cite{Filho99}, which looked at the fraction of votes received by
candidates for the positions of state deputies.  The resulting frequency
distribution was fit by a power law with exponent very close to $-1$. The
cumulative distribution, however, revealed that about $90 \%$ of the
candidates' votes followed a log-normal distribution, with a large
dispersion that resulted in the apparent power law.   
\begin{figure*}
\includegraphics[width=.4\linewidth]{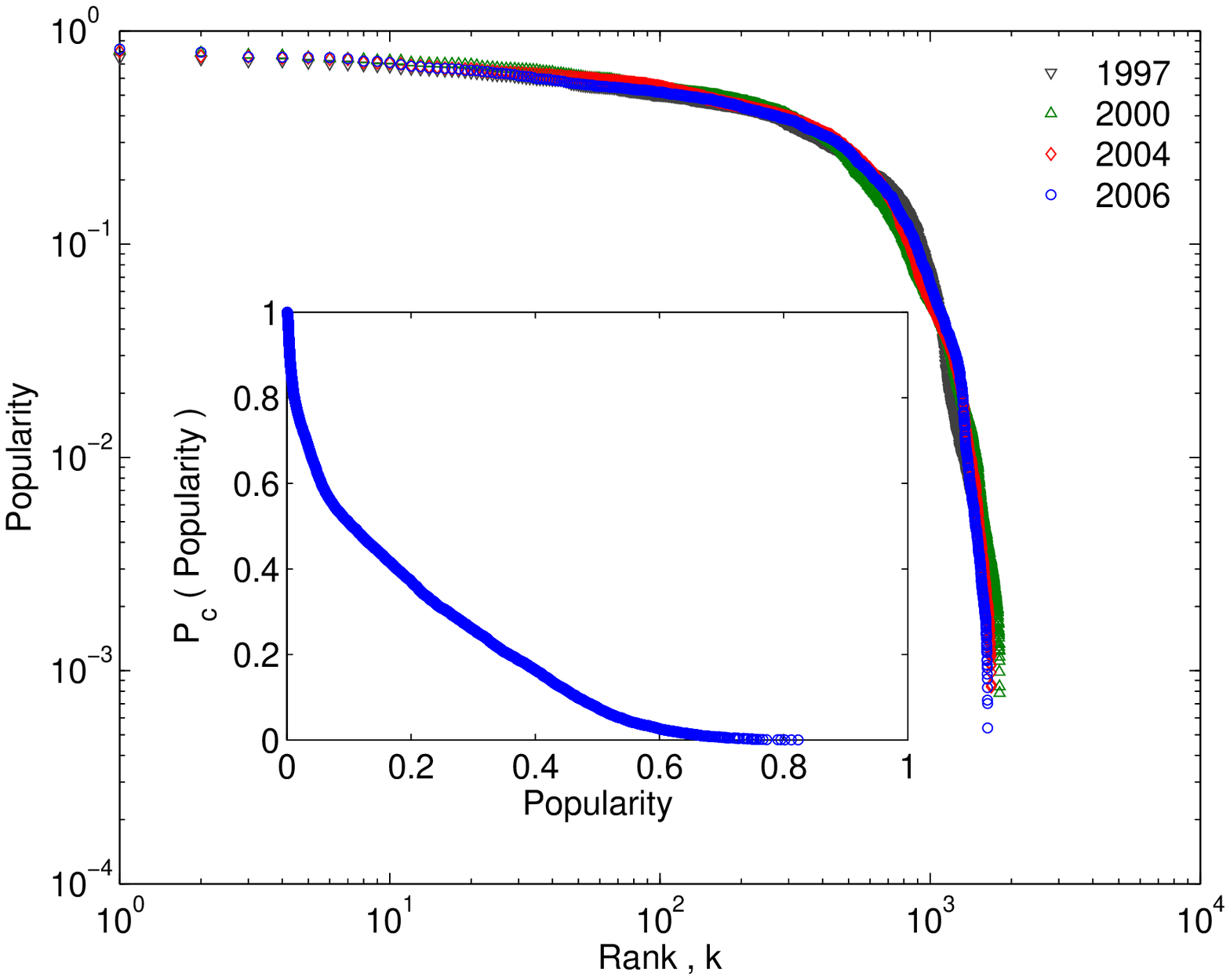}
\includegraphics[width=.4\linewidth]{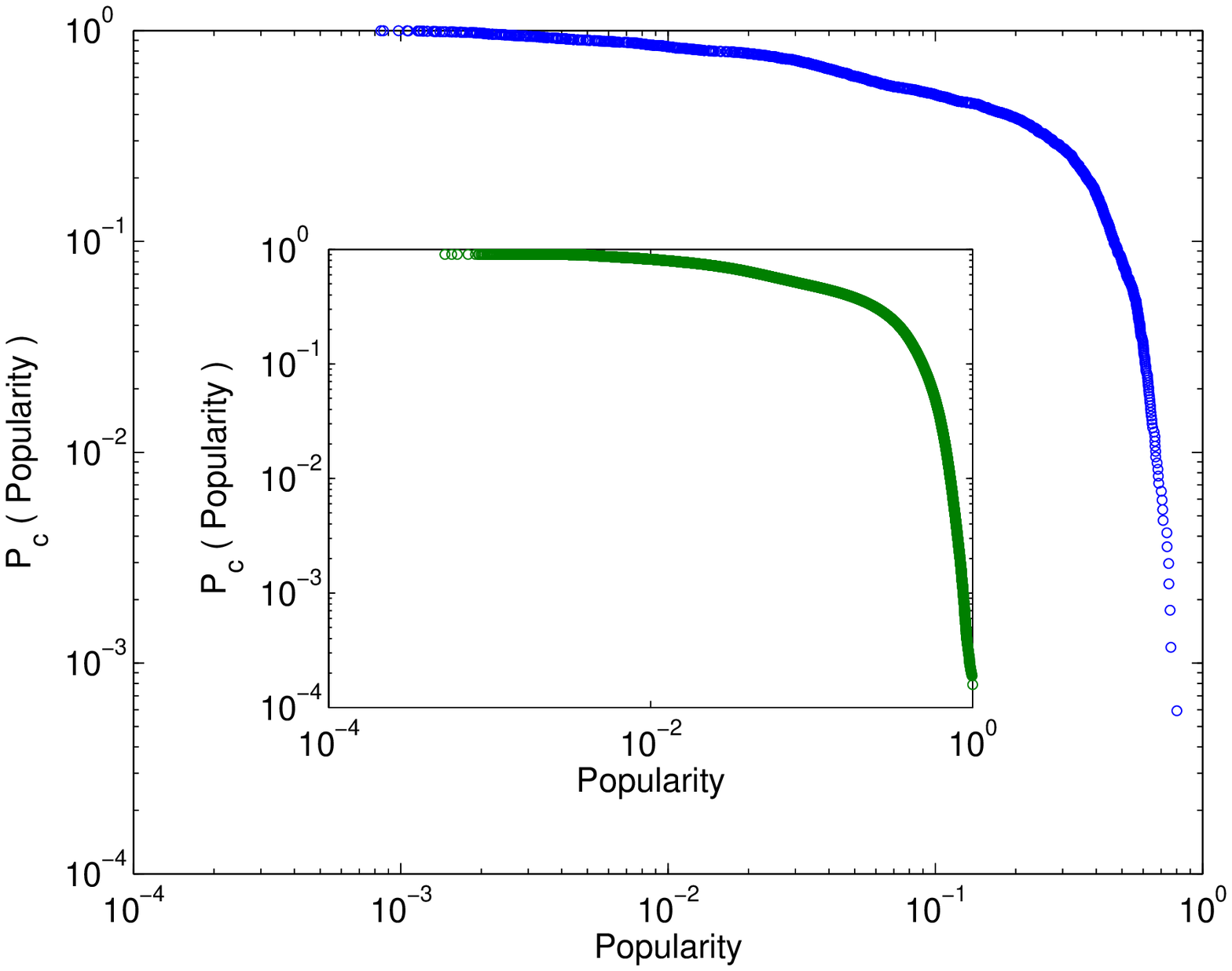}
\caption{Canadian elections:
(Left) The rank-ordered	plot of candidate popularity measured by the 
fraction of votes received by him or her, for four successive general 
elections. 
The inset shows the cumulative frequency distribution function for this
popularity measure. Note the region of linear decay in the middle of the
curve.
(Right) Cumulative probability distribution function for the fraction of votes
received by a candidate for all constituencies in the 2000 general election. 
The inset shows the cumulative distribution function of the vote fraction
for candidates for all polling booths at each constituency in the above 
election. Note that a constituency can have hundreds of polling booths.}
\label{ss:canada}
\end{figure*}

We have carried out an analysis of the distribution of votes for a number
of general elections in Canada and India.  The data about votes for
individual candidates in Canada was obtained from the website Elections
Canada On-line~\footnote{http://www.elections.ca/} for the general
elections held in 1997, 2000, 2004 and 2006. The total number of candidates
in each election varied between 1600-1800, there were over $\sim 300$
electoral constituencies and the total number of votes cast varied around
13 million.  Each constituency was divided into hundreds of polling
stations, thereby allowing us to obtain a micro-level picture of the
popularity of the candidates at a particular constituency across the
different polling stations. Fig.~\ref{ss:canada}~(left) shows the results
of our analysis, indicating an exponential decay of the tail of the
popularity distribution for all the elections being considered.  The
results don't change even if we consider the number of votes, rather than
the vote fraction. Fig. \ref{ss:canada}~(right) shows that the distribution
of popularity across polling stations has almost an identical distribution
to that seen over the larger scale of electoral constituencies.  Note that
we did not observe the popularity of parties for Canada, as the total
number of parties were only about 10.

\begin{figure*}
\includegraphics[width=.4\linewidth]{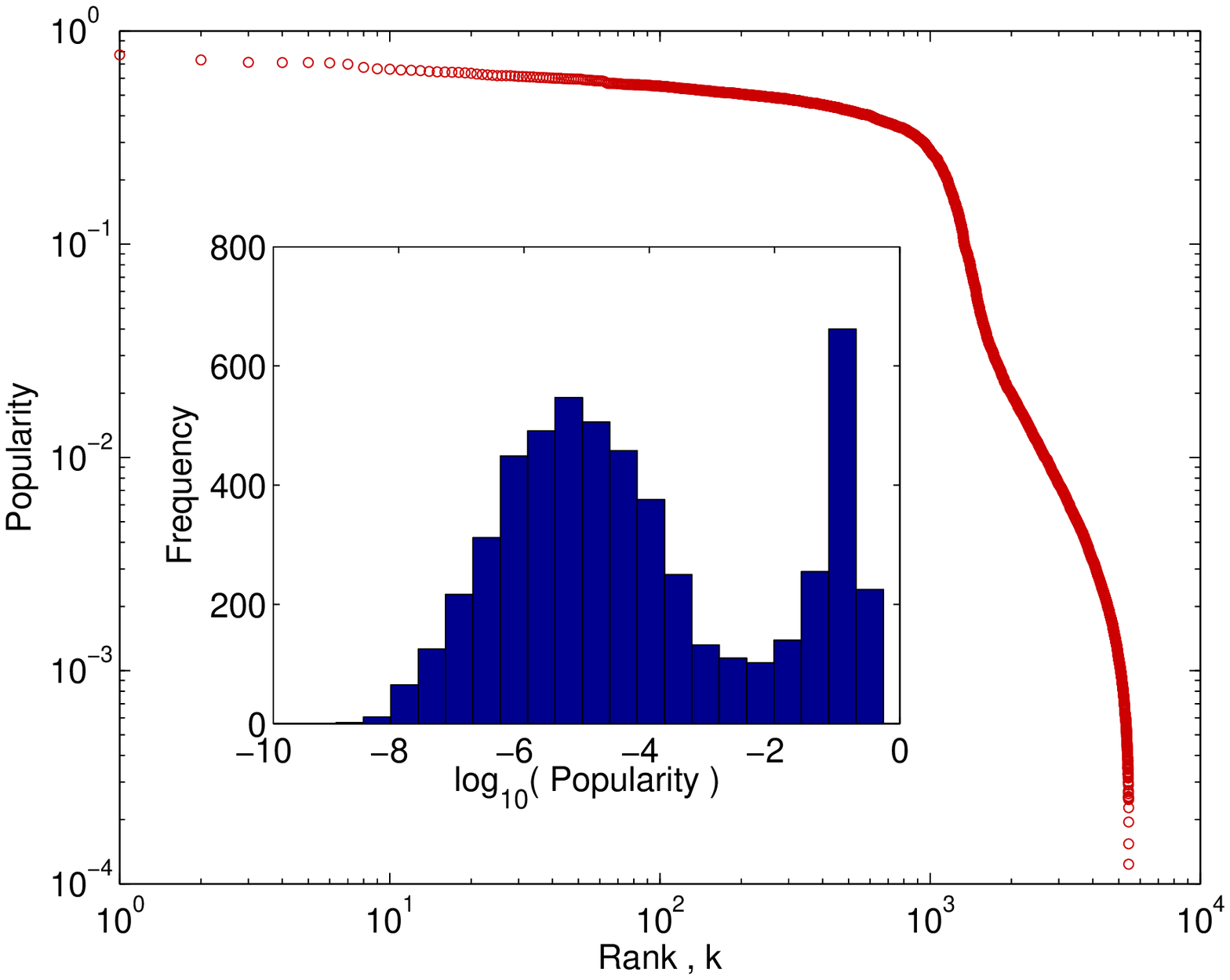}
\includegraphics[width=.4\linewidth]{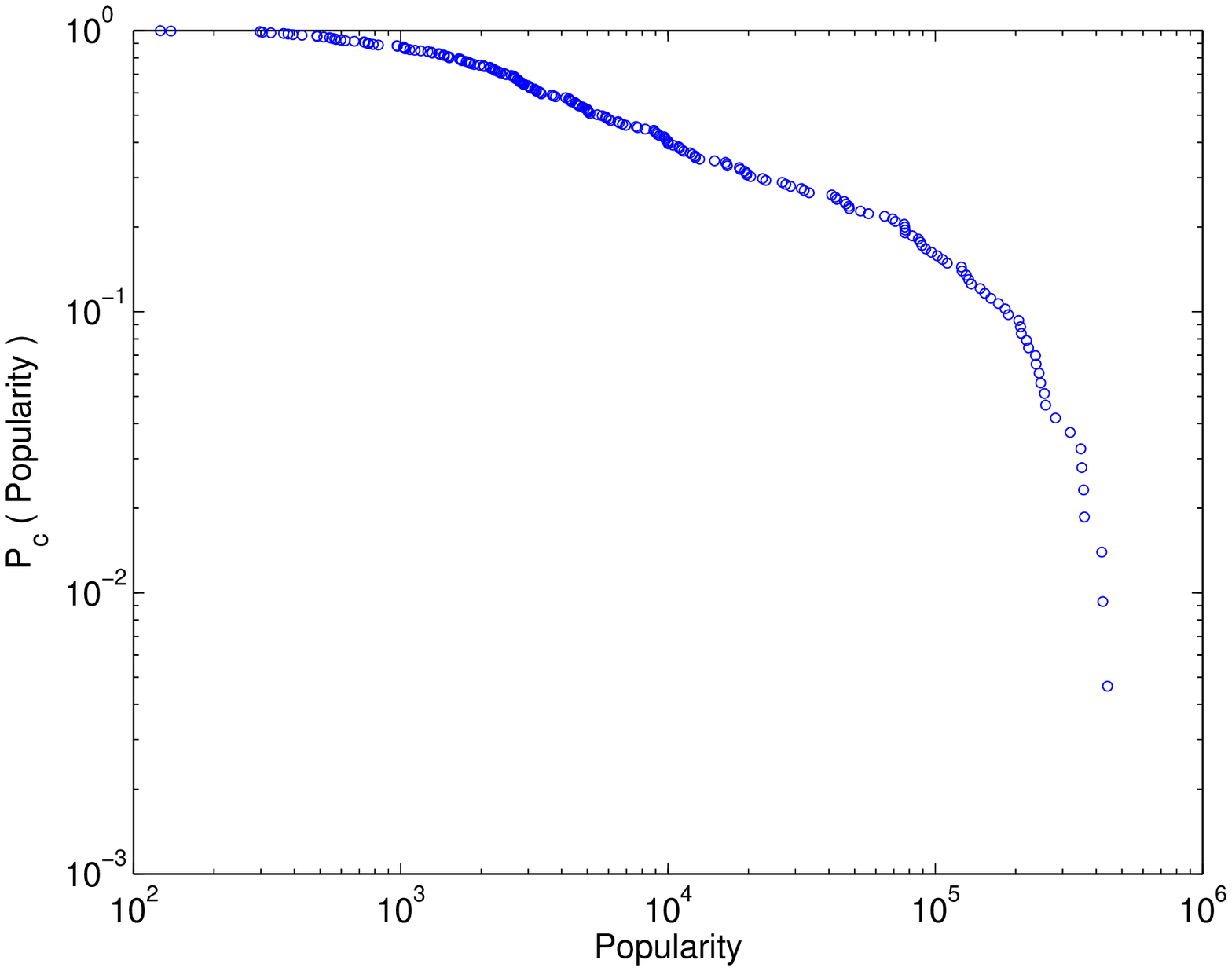}
\caption{Indian election:
(Left) The rank-ordered plot of candidate popularity measured by the fraction 
of votes received, for the 2004 Lok Sabha election. The inset shows the
frequency distribution of the vote fraction, clearly indicating a bimodal
nature with candidates receiving either most of the votes cast or very few.
(Right) Cumulative probability distribution function of party popularity for 
the 2004 election, measured by the fraction of votes received by candidates 
from that party, over all the constituencies it contested in.}
\label{ss:india}
\end{figure*}
Next, we looked at the corresponding data for the 2004 general elections in
India obtained from the website of the Election Commission of
India~\footnote{http://www.eci.gov.in/}. The total number of candidates is
5435, about half of whom belonged to 230 registered parties, who contested
from a total of 543 electoral constituencies, while the total number of
votes cast was about 400 million.  Fig.~\ref{ss:india}~(left) shows that
the rank-ordered popularity (measured by the vote fraction) distribution
for candidates in an Indian general election is qualitatively similar to
that of Canada, except for the presence of a kink indicative of the bimodal
nature of the distribution. This implies that candidates either receive
most of the votes cast by electors in that constituency or very few votes.
It maybe due to the very large number of independent candidates (i.e.,
without affiliation to any recognized party) in Indian elections compared
to Canada. This is supported by our analysis of popularity of recognized
political parties [Fig.~\ref{ss:india}~(right)] that shows an exponential
decay at the tail. Note that the popularity of a party is measured by the
total votes received by a party divided by the number of constituencies in
which it contested. This is same (upto a scaling constant) as the
percentage of votes received by candidates belonging to a party, averaged
over all the constituencies in which the party had fielded candidates. 

\subsubsection{Books.}
\begin{figure*}
\includegraphics[width=.4\linewidth]{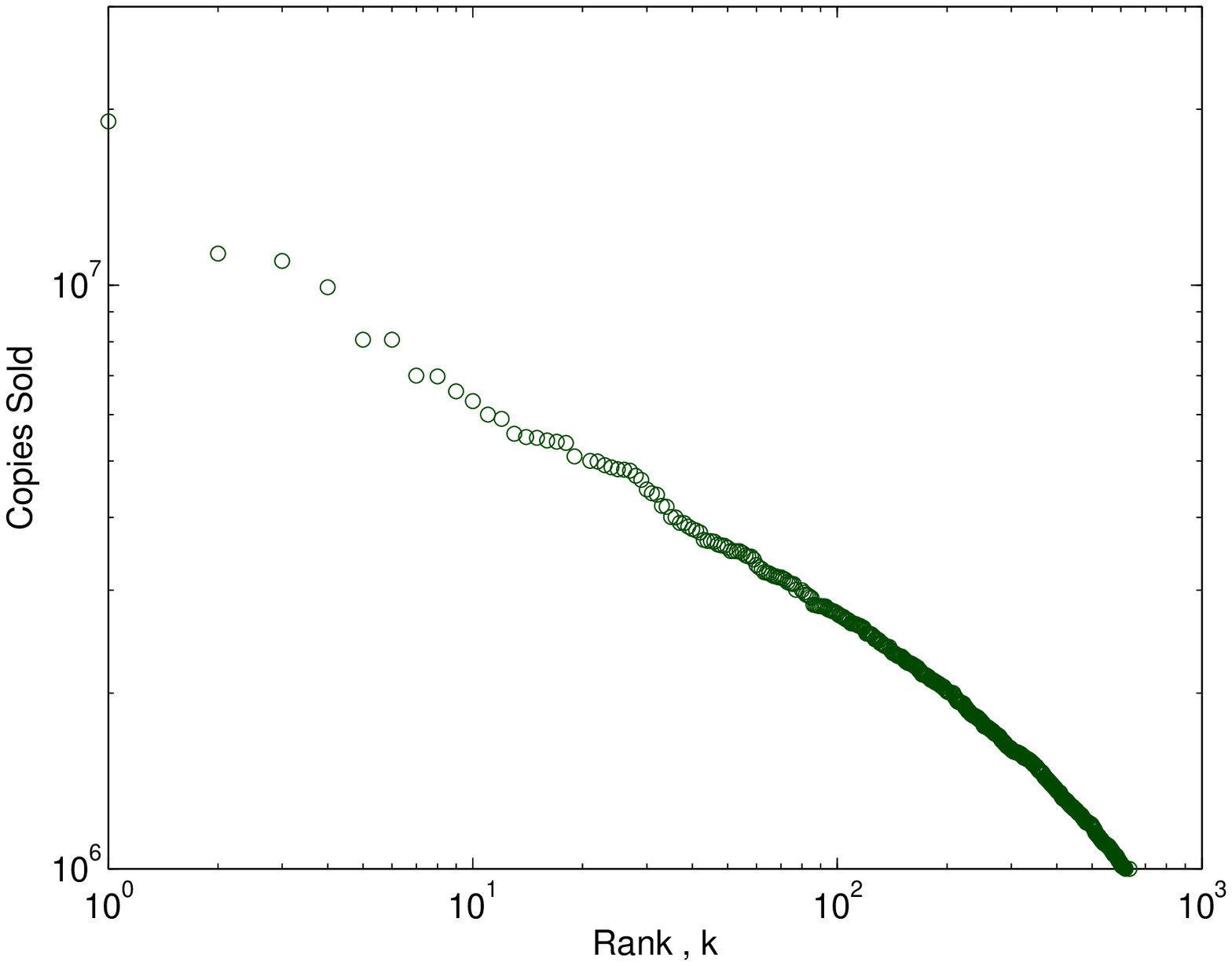}
\includegraphics[width=.4\linewidth]{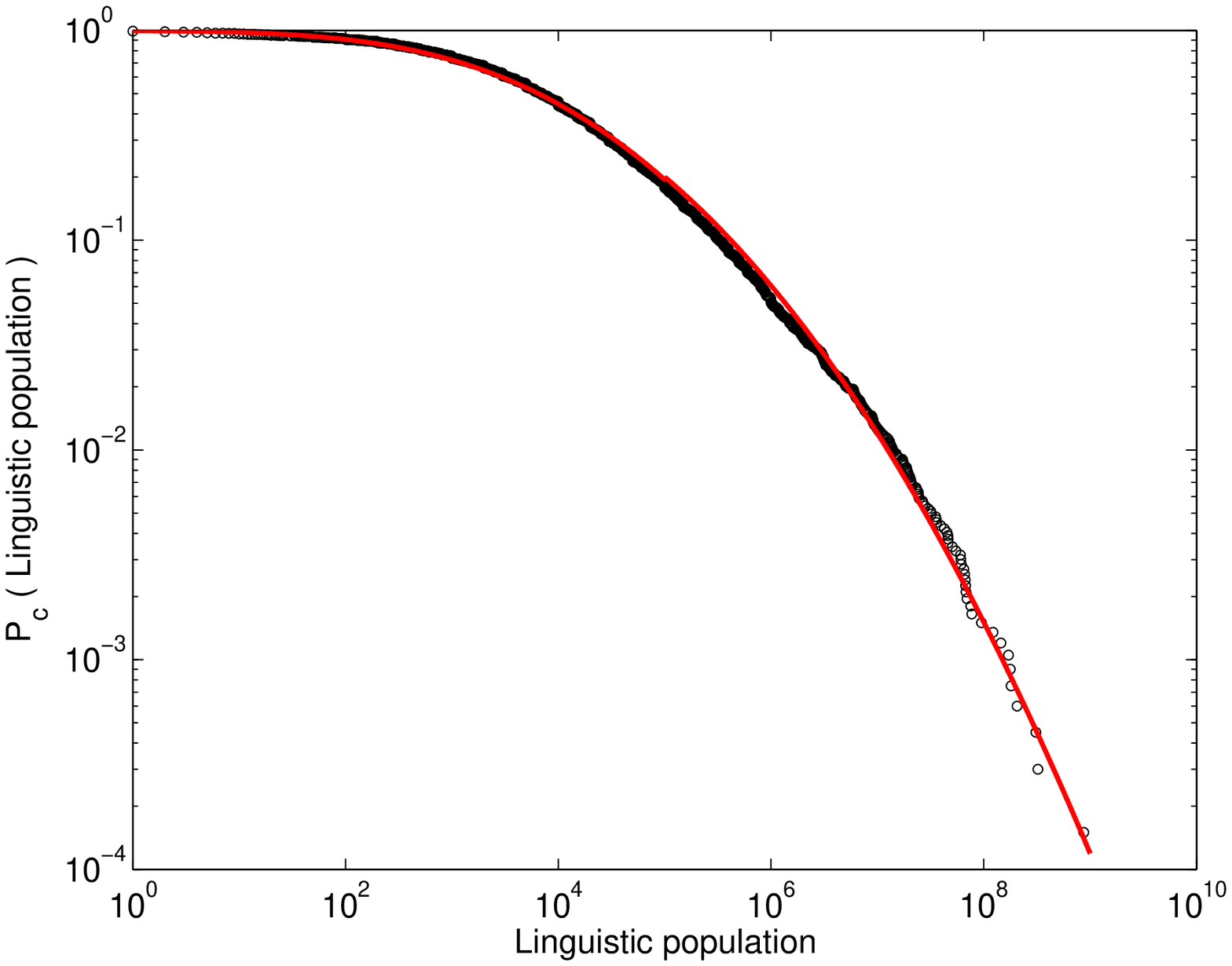}
\caption{(Left) The rank-ordered plot of bestselling books (that sold 2
million copies or more) according to the number of copies sold in USA
between 1895 to 1965.  Adapted from Ref.~\cite{Newman05}, data provided by
M.~E.~J. Newman.  (Right) Cumulative distribution function for the size of
the population of first-language speakers for over 6650 languages. The data
was obtained from {\em Ethnologue}. The curve ({\em in red}) indicates the
best log-normal fit to the data.}
\label{ss:booklanguage}
\end{figure*}

An obvious popularity distribution based on product sales is that of books,
especially in view of the record-breaking sales in recent times of the {\em
Harry Potter} series of books. However, the lack of freely available data
about exact sales figures has so far prevented detailed analysis of book
popularity. It was reported in a recent paper~\cite{Sornette04}, that the
cumulative distribution of book sales from the online bookseller
Amazon~\footnote{http://www.amazon.com} has a power-law tail with $\alpha
\sim 2$. However, one should note that Amazon does not reveal exact sales
figures, but rather only the rank according to sales; therefore, this
distribution was actually based on a heuristic relation between rank and
sales proposed by Rosenthal~\cite{Rosenthal06}. Needless to say, this is at
best a very rough guide to the exact sales figures (e.g., although the sale
of {\em Harry Potter and the Half-Blood Prince} fluctuated a lot during the
few weeks following its publication, it remained steady as the top ranked
book in Amazon) and is likely to yield misleading distribution of sales.  A
more reliable dataset, if somewhat old, has been compiled by
Hackett~\cite{Hackett67} for the total number of copies sold in USA of the
top 633 bestselling books between 1895 and 1965.  Newman~\cite{Newman05}
has reported the maximum likelihood estimate for the exponent of the power
law fit to this data as $\alpha \sim 2.51$.
Fig.~\ref{ss:booklanguage}~(left) shows the rank-ordered plot of this data,
indicating an approximate power law fit for slightly more than a decade,
with an exponent of $-0.4$. 

\subsubsection{Language.}
Fig.~\ref{ss:booklanguage}~(right) shows the cumulative distribution of the
first-language speaker population for different languages around the world.
The data has been obtained from {\em
Ethnologue}~\footnote{http://www.ethnologue.com/} which provides the number
of first-language speakers (over all countries in the world) wherever
possible.  Out of a total of 7299 languages listed in its 15th edition, we
have considered above 6650 languages for which information about the number
of speakers is available.  The figure shows a long tail with an
approximately log-normal fit; maximum likelihood estimates of parameters
for the corresponding distribution are $\mu=8.78$ and $\sigma=3.17$. Note
that this kind of popularity distribution is different from the others we
have discussed so far as the speakers are not really free to choose their
first language; rather this is connected to the population growth rate of a
particular linguistic community. A similar kind of popularity distribution
is that for family names, which has been analysed by Miyazima et
al~\cite{Miyazima99} for Japanese family names and Newman~\cite{Newman05}
for American family names, both reporting cumulative distribution functions
with power law tails having $\alpha$ close to 1.  However, for Korean
family names~\cite{Kim04} the distribution was reported to be exponentially
decaying. 

\subsubsection{Other Popularity distributions.}
Unlike the distribution of family names discussed above, the frequency of
occurrence of given names (or first names) are indeed subject to waves of
popularity, with certain names appearing to be very common at a particular
period. A recent study~\cite{Galbi05} has looked at the distribution of
most popular given names in England and Wales over the past millennium, and
has claimed a long-tailed distribution for the same. 
Another popularity distribution is that of tourist destinations, as
measured by the number of tourist arrivals over a time period. 
A study~\cite{Hazari04} that has ranked 89 countries, focussing on the
period 1980-1990, have found evidence for a log-normal distribution as
the best fit to the data. 

The occurrence of superstars (i.e., extremely successful performers) in
popular music has led to a relatively large amount of literature by
economists on the occurrence of
popularity~\cite{Rosen81,Adler85,MacDonald88,Hamlen94}.  Chung \& Cox have
used the number of gold-records by performers as the measure of their
artistic success, and found the tail of this popularity distribution to
approximately follow a power law~\cite{Chung94}.  Another
study~\cite{Davies02} looked at the longevity of music bands in the list of
Top 75 best-selling recordings, and observed a stretched exponential
distribution~\footnote{While the term stretched exponential distribution is
quite common in the physics literature, we observe that in other scientific
fields it is more commonly referred to as Weibull distribution.}.  However,
a more recent study~\cite{Giles05} has shown the survival probability of a
music recording on the {\em Billboard} Hot 100 chart to be fit better by
the log-logistic distribution.  

\subsection{Time-evolution of popularity}
\begin{figure}
\includegraphics[width=.85\linewidth]{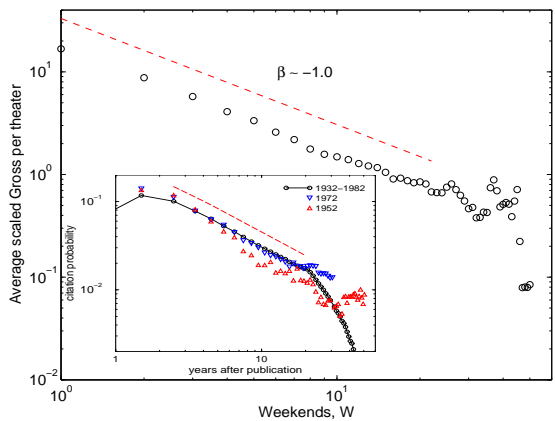}
\caption{ Weekend gross per theater for a movie (scaled by the average
weekend gross over its theatrical lifespan), after it has run for $W$
weekends, averaged over the number of movies that ran for that long.  The
initial decline follows a power-law with exponent $\beta \simeq -1$ (the
fit is shown by the broken line).  The inset (from Ref.~\cite{Redner04})
shows the probability that a paper will be cited $t$ years after
publication in a Physical Review journal, in the years 1952 and 1972, as
well as over the period 1932-1982. Over the range of $2-20$ years the
integrated data is consistent with a power law decay having an exponent
$-0.94$ ({\em broken line in red}).}
\label{ss:moviedynamics}
\end{figure}
Here we look briefly at how popularity evolves over time. For movies, we
look at the gross income per theater over time
(Fig.~\ref{ss:moviedynamics}).  This is a better measure of the dynamics of
movie popularity than the time-evolution of the weekly overall gross
income, because a movie that is being shown in a large number of theaters
has a bigger income simply on account of higher accessibility for the
potential audience. Unlike the overall gross that decays exponentially with
time, the gross per theater shows a power-law decay in time with exponent
$\beta \simeq -1$~\cite{Sinha05}.  This has a striking similarity with the
time-evolution of popularity for scientific papers in terms of citations.
It has been reported that the citation probability to a paper published $t$
years ago, decays approximately as $1/t$~\cite{Redner04}
[Fig.~\ref{ss:moviedynamics}~(inset)]. Note that, Price~\cite{Price76} had
also noted a similar behavior for the decay of citations to papers listed
in the Science Citation Index.  In a very different context, namely, the
decay in the popularity of a website (as measured by the rate of download
of papers from the site) over time $t$ has also been reported to follow an
inverse power-law, but with a different exponent~\cite{Sornette00}.

\subsection{Discussion}
The selection of (mostly) long-tailed empirical popularity distributions
presented above underlines the following broad features of such
distributions: (i) the entire distribution seem to be fit
by a log-normal curve (in the few cases where the entire 
distribution is not available, the upper tail seems to fit a power law with
characteristic exponent $\alpha $ which is often close to $1$, corresponding 
to the exact form of Zipf's law); (ii) in some cases the distribution shows a
bimodal character, with most of the instances occurring at the two
ends of the distribution; (iii) the decay of popularity in some cases
seem to show a simple power law decay, declining inversely with time
elapsed since release; (iv) the persistence time at high levels of
popularity show a Weibull distribution in many instances. 

The first of these features may come somewhat as a surprise, because
for many popularity distributions, power law tails have been reported
with various exponents, often significantly different from 1. However,
we observe that very often log-normal distributions have been
mistakenly identified as having power law tails.  In fact this is a
very common error, especially if the variance of the log-normal
distribution is sufficiently large. To see this, note that the
log-normal distribution, 
\begin{equation} 
P(x) = \frac{1}{x \sigma \sqrt{2 \pi}} e^{- ({\rm ln} x - \mu)^2/2 {\sigma}^2},
\label{ss:lognormal}
\end{equation}
can be written as (on taking logarithm on both sides),
\begin{equation}
{\rm ln} P(x) = -\frac{({\rm ln} x )^2}{2 \sigma^2} + (\frac{\mu}{\sigma^2} -1)
- {\rm ln} \sqrt{2 \pi} \sigma - \frac{\mu^2}{2 \sigma^2},
\end{equation}
which is a quadratic curve in a doubly logarithmic plot. However, a
sufficiently small part of the curve will appear as a straight line,
with the slope depending on which segment of the curve one is
focussing attention~\cite{Newman05,Mitzenmacher03}.This is the
origin of most of the power law tails with exponent $\alpha \neq 1$
that has been reported in the literature on popularity distributions.

\section{Models of Popularity Distribution}
\begin{figure}
\includegraphics[width=.95\linewidth]{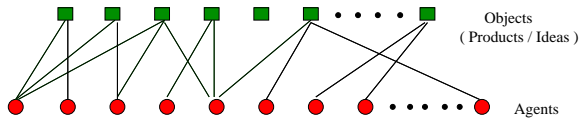}
\caption{A schematic diagram of the emergence of popularity as a relation
between agents and objects (products or ideas).}
\label{ss:modelschematic}
\end{figure}

From the perspective of physics, popularity can be viewed as an emergent
outcome of the collective decision process in a society of individual
agents exercising their free will (as reflected in their individual
preferences) to choose between alternative products or ideas
(Fig.~\ref{ss:modelschematic}).  In a system without authoritarian control,
agents differ in their personal preferences which are determined by the
information available to the agent about the possible alternatives.
However, in any real-life scenario with uneven access to information, a
seemingly well-informed agent may influence the choice of several other
agents~\cite{Bikchandani92}.  Thus, the emergence of a popular product is a
result of the self-organized coordination of choices made by heterogeneous
entities.

The simplest model of collective choice is one where the agents decide
independently of each other and select alternatives at random with a
one-step decision process. It is easy to see that the possible alternatives
will not be significantly different in terms of popularity from each other.
In particular, the popularity distribution arising from such a process will
not have long tails.  There are two possible alternative modifications of
this simple model that will allow it to generate distributions similar to
the ones seen empirically. The first option is to allow interactions
between agents where the choice of one agent can influence that of another.
While this is often true in real-life, we also observe long-tailed
distributions much before the interaction among agents (and the resulting
dissemination of information) has had a chance to influence the popularity.
For example, the long-tailed distribution of movie popularity, in terms of
gross earning, is seen at the opening weekend itself, long before potential
movie viewers have had a chance to be influenced by other moviegoers. The
second option for generating realistic popularity distribution gets around
this problem: here we replace the single-step decision process by one
comprising of multiple sub-decisions (as there may be many factors involved
in making a particular decision), each of which contribute to the overall
decision to purchase a particular product.  Therefore, the probability of
any particular entity achieving a particular degree of popularity can be
expressed as the product of probabilities of each of the underlying factors
satisfying the required condition to make an agent opt for that entity. As
is easily seen, the resultant distribution arising from such a
multiplicative stochastic process has a log-normal form, agreeing with many
of the empirically observed distributions~\footnote{One can argue that the
probability distribution of collective choice may also reflect the
distribution of quality amongst various competing entities; however, in
this case the popularity distribution would be essentially identical to the
quality distribution, which a priori can follow any arbitrary distribution.
The universality of long-tailed popularity distributions and the seeming
absence of any correlation between popularity and quality (when it can be
measured in any well-defined manner) would argue against this hypothesis.}.

While the bulk of the popularity distributions, showing a log-normal
nature, can therefore be plausibly explained as the product of the
multiplicative stochastic structure underlying even apparently simple
decision processes, this would still leave unanswered the reason for the
wide occurrence of Zipf's law in other instances.  We now turn to the first
option for extending the simple model outlined above, i.e., investigating
the influence of an agent's choice behavior on other agents.  It turns out
there have been many proposed mechanisms to explain the ubiquity of
power-law tailed distributions employing interactions.  However, from the
point of view of the present paper, the most relevant (and general) model
seems to be the Yule process~\cite{Yule25}, as modified by
Simon~\cite{Simon55}. This is essentially a cumulative advantage process by
which the relatively more popular entities get even more popular by virtue
of being more well-known.

The Yule-Simon process can be described as follows: Suppose initially there
are $n$ agents, each of whom are free to choose one of a number of
products. Subsequently, the number of agents is augmented by unity at each
time step. At any point in time, when the total number of agents is $m$,
the number of distinct products, each of which have been chosen by $k$
agents is denoted by $f(k,m)$.  Then, given that, (i) there is a constant
probability, $\gamma$, that an agent chooses a completely new product
(i.e., one that has not been chosen before by any of the agents) and (ii)
the probability of choosing a product that has already been chosen by $k$
agents is proportional to $kf(k,n)$, one obtains an asymptotic popularity
distribution that has a power-law tail~\footnote{Note that, the models of
Price~\cite{Price76}, Barabasi-Albert~\cite{Barabasi99} and
Redner~\cite{Redner02} are all special cases of this general mechanism.}
with exponent $\alpha = \frac{1}{1-\gamma}$.  If the appearance of a new
product is relatively infrequent, i.e., $\gamma$ is extremely small, then
the exponent $\alpha \simeq 1$ (i.e., Zipf's law).

Another feature of popularity distributions that has been mentioned earlier
is that, in some cases, they appear to have a bimodal nature. We now
present a simple agent-based model~\cite{Sinha04b} that shows how bimodal
and unimodal distributions of popularity can arise very simply through
agents interacting with each other, and reacting to information about what
the majority are choosing in the previous time step.

\subsection{A Model for Bimodal Distribution of Collective Choice}
We have already discussed the simplest model of collective choice in which
individual agents make completely independent decisions.  For binary choice
(i.e., each agent can only choose between two options) the emergence of
collective choice is equivalent to a one-dimensional random walk with the
number of steps equal to the number of agents. Therefore, the outcome will
be normally distributed, with the most probable outcome being an equal
number of agents choosing each alternative.  While such unimodal
distributions of popularity are indeed observed in some situations, as
mentioned earlier in this article many real-life examples show the
occurrence of bimodal distributions indicative of highly polarized choice
behavior among agents resulting in the emergence of a highly popular
product.  This polarization suggests that agents not only opt for certain
choices based on their personal preferences, but are also influenced by
other agents in their social neighborhood.  Also, the personal preferences
may themselves change over time as a result of the outcome of previous
choices, e.g., whether or not their choice agreed with that of the
majority. This latter effect is an example of global feedback process that
we think is crucial in the occurrence of bimodal behavior.

We now present a general model of collective decision that shows how
polarization in the presence of individual choice volatility can be
achieved with an adaptation and learning dynamics of the personal
preference.  In this model, the choice of individual agents are not only
affected by those of their neighbors, but, in addition, their preference is
modified by their previous choice as well as information about how
successful their previous choice behavior was in coordinating with that of
the majority.  Here it is assumed that information about the intrinsic
quality of the alternative products is inaccessible to the agent, who takes
the cue from what the majority is choosing to decide which one is the
``better choice''.  Examples of such limited global information about the
majority's preference available to an agent are the results of consumer
surveys and publicity campaigns disseminated through the mass media.

The simplest, binary choice version of our model is defined as follows.
Consider a population of $N$ agents, each of whom can be in one of two
choice states $S = \pm 1$ (e.g., to buy or not to buy a certain product, to
vote Party A or Party B, etc.). In addition, each agent has an individual
preference, $\theta$, that is chosen from a uniform random distribution
initially.  At each time step, every agent considers the average choice of
its neighbors at the previous instant, and if this exceeds its personal
preference, makes the same choice; otherwise, it makes the opposite choice.
Then, for the $i$-th agent, the choice dynamics is described by:
\begin{equation}
S_i^{t+1} = {\rm sign} ( \sum_{j \in {\cal N}} J_{ij} S_j^t - \theta_i^t),
\label{ss:modeleq1}
\end{equation}
where sign ($x$) = $+1$, if $x > 0$, and = $-1$, otherwise.  The coupling
coefficient among agents, $J_{ij}$, is assumed to be a constant ($= 1$) for
simplicity and normalized by $z$ ($= |{\cal N}|$), the number of neighbors.
In a lattice, ${\cal N}$ is the set of spatial nearest neighbors and $z$ is
the coordination number, while in the mean field approximation, ${\cal N}$
is the set of all other agents in the system and $z = N - 1$. 
\begin{figure*}
\includegraphics[width=.4\linewidth]{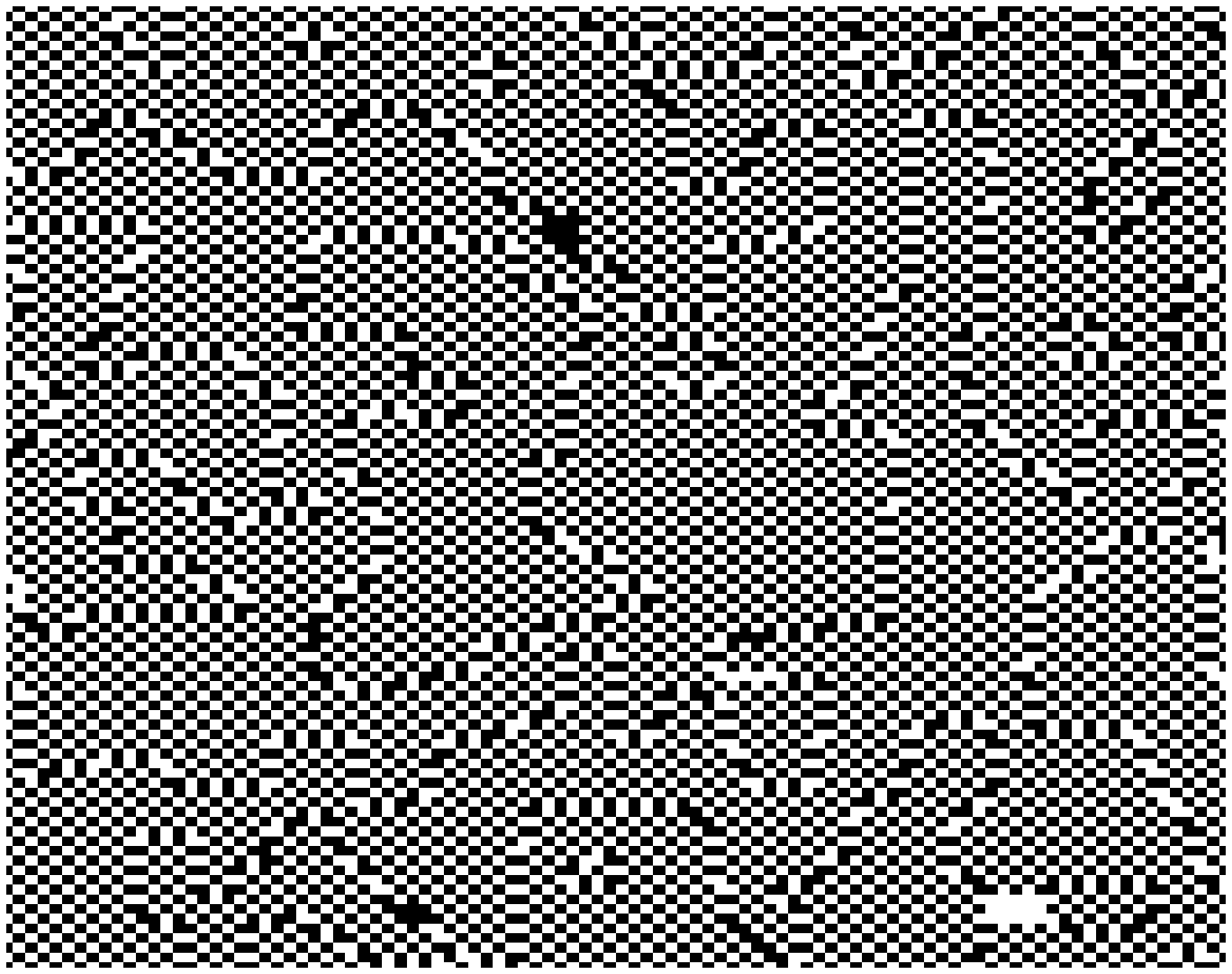}
\includegraphics[width=.4\linewidth]{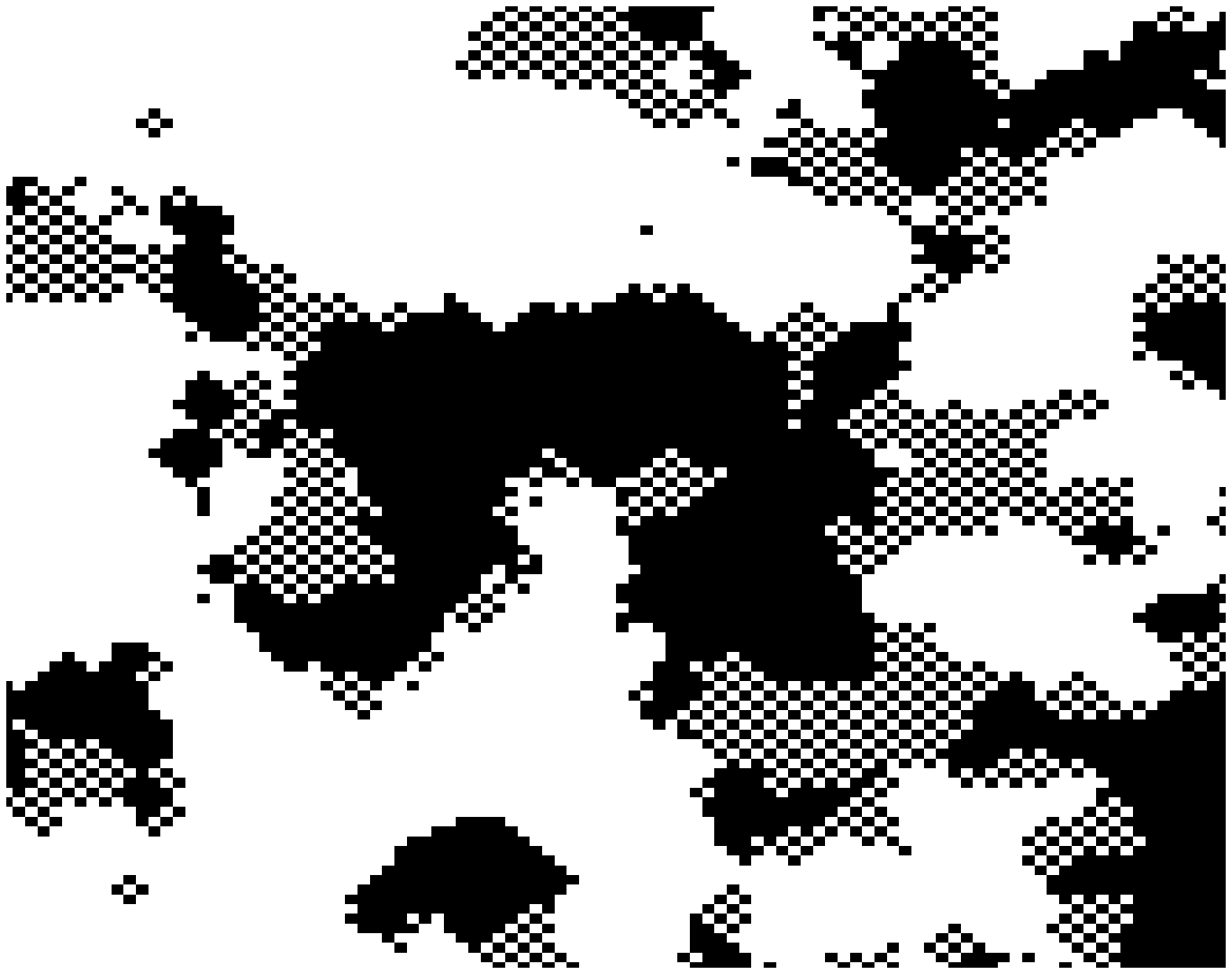}
\caption{(Left) The spatial pattern of choice ($S$) in the absence of
learning ($\lambda = 0$) in a two-dimensional square lattice of $1000
\times 1000$ agents after 500 iterations starting from a random
configuration. The figure is a magnified view of the the central $100
\times 100$ region showing the absence of long-range correlation among the
agents.  (Right) The spatial pattern of choice ($S$) with learning
($\lambda = 0.05$) in the same system, with a majority of agents now in the
choice state $S = +1$. The magnified view of the central $100 \times 100$
region shows coarsening of regions having agents aligned in the same choice
state.}
\label{ss:modelspatial}
\end{figure*}

The individual preference, $\theta$, evolves over time as:
\begin{eqnarray}
\theta_i^{t+1} & = & \theta_i^t + \mu S_i^{t+1} + \lambda S_i^t, ~{\rm if}~
S_i^{t} \neq {\rm sign} (M^t), \nonumber \\
& = &   \theta_i^t + \mu S_i^{t+1}, {\rm otherwise}, 
\label{ss:modeleq2}
\end{eqnarray}
where $M^t = (1/N) \sum_j S_j^t$ is the collective decision of the entire
community at time $t$.  Adjustment to previous choice is governed by the
adaptation rate $\mu$ in the second term on the right-hand side of
Eq.~(\ref{ss:modeleq2}), while the third term, governed by the learning
rate $\lambda$, represents the correction when the individual choice does
not agree with that of the majority at the previous instant. The
desirability of a particular choice is assumed to be related to the
fraction of the community choosing it; hence, at any given time, every
agent is trying to coordinate its choice with that of the majority. Note
that, for $\mu = 0, \lambda = 0$, the model reduces to the well-known
zero-temperature, random field Ising model (RFIM).

{\em Random neighbor and mean field model.}
For mathematical convenience, we choose the $z$ neighbors of an agent at
random from the $N-1$ other agents in the system. We also assume this
randomness to be ``annealed'', i.e., the next time the same agent interacts
with $z$ other agents, they are chosen at random anew.  Thus, by ignoring
spatial correlations, a mean field approximation is achieved. 

For $z = N - 1$, i.e., when every agent has the information about the
entire system, it is easy to see that, in the absence of learning ($\lambda
= 0$), the collective decision $M$ follows the evolution equation rule:
$M^{t+1} = {\rm sign} [ (1-\mu) M^t - \mu \sum_{\tau = 1}^{t-1}
M^{\tau}].$ For $0 < \mu < 1$, the system alternates between the ordered
states $M=\pm 1$ with a period $\sim 4/\mu$. The residence time at any one
state ($\sim 2/\mu$) diverges with decreasing $\mu$, and for $\mu = 0$, the
system remains fixed at one of the ordered states corresponding to $M=\pm
1$, as expected from RFIM results.  At $\mu = 1$, the system remains in the
disordered state, so that $M = 0$.  Therefore, we see a transition from a
bimodal distribution of the collective decision, $M$, with peaks at
non-zero values, to an unimodal distribution of $M$ centered about 0, at
$\mu_c = 1$. When we introduce learning, so that $\lambda > 0$, the agents
try to coordinate with each other and at the limit $\lambda \rightarrow
\infty$ it is easy to see that $S_i = {\rm sign}(M)$ for all $i$, so that
all the agents make identical choice.In the simulations, we note that the
bimodal distribution is recovered for $\mu =1$ when $\lambda \geq 1$. 

For finite values of $z$, the population is no longer ``well-mixed'' and
the mean-field approximation becomes less accurate the lower $z$ is.  For
$z < < N$, the critical value of $\mu$ at which the transition from a
bimodal to a unimodal distribution occurs in the absence of learning,
$\mu_c < 1$. For example, $\mu_c = 0$ for $z = 2$, while it is 3/4 for $z =
4$. As $z$ increases $\mu_c$ quickly converges to the mean-field value,
$\mu_c = 1$.  On introducing learning ($\lambda > 0$) for $\mu > \mu_c$, we
again notice a transition to an ordered state, with more and more agents
coordinating their choice. 

$Lattice.$ To implement the model when the neighbors are spatially related,
we consider $d$-dimensional lattices ($d = 1, 2, 3$) and study the dynamics
numerically. We report results obtained in systems with absorbing boundary
conditions; using periodic boundary conditions leads to minor changes but
the overall qualitative results remain the same.  It is worth noting that
the adaptation term disrupts the ordering expected from results of the RFIM
for $d = 3$, so that for any non-zero $\mu$ the system is in a disordered
state when $\lambda = 0$. 

\begin{figure*}
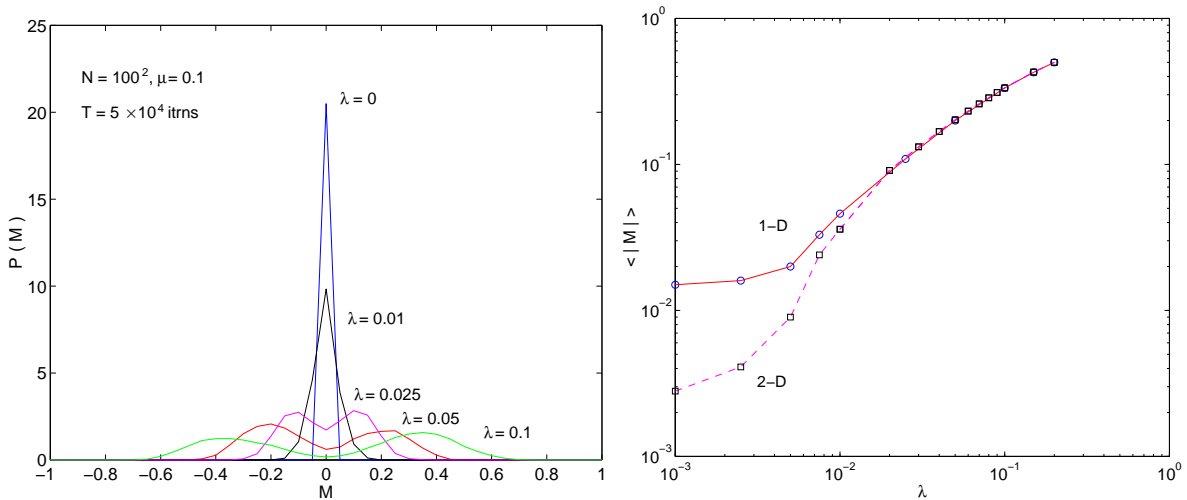

\includegraphics[width=.45\linewidth]{sinha-model_distrn.eps}
\includegraphics[width=.42\linewidth]{sinha-model_orderparameter.eps}
\caption{(Left) The probability distribution of the collective decision $M$
in a two-dimensional square lattice of $100 \times 100$ agents. The
adaptation rate $\mu = 0.1$, and the learning rate $\lambda$ is increased
from 0 to 0.1 to show the transition from unimodal to bimodal behavior.
The system was simulated for $5 \times 10^4$ iterations to obtain the
distribution.  (Right) The order parameter $< | M | >$ for one- and
two-dimensional lattices. The adaptation rate is $\mu = 0.1$, while
$\lambda$ is increased gradually to show the transition to an ordered
state. Note that for higher values of $\mu$ the two curves are virtually
identical. There is very little system size-dependence of the curves.}
\label{ss:modeldistrn}
\end{figure*}
In the absence of learning ($\lambda = 0$), starting from a initial random
distribution of choices and personal preferences, we observe only very
small clusters of similar choice behavior
[Fig.~\ref{ss:modelspatial}~(left)] and the average choice $M$ fluctuates
around 0. In other words, at any given time an equal number (on average) of
agents have opposite choice preferences.  Introduction of learning in the
model ($\lambda > 0$) gives rise to significant clustering as well as a
non-zero value for the collective choice $M$. We find that the probability
distribution of $M$ [Fig.~\ref{ss:modeldistrn}~(left)] evolves from a
single peak at 0, to a bimodal distribution as $\lambda$ increases from 0.
This is similar to second-order phase transition in systems undergoing
qualitative changes at a critical threshold. The collective decision $M$
switches periodically from a positive value to a negative value having an
average residence time which diverges with $\lambda$ and with $N$.  For
$\mu > \lambda > 0$, large clusters of agents with identical choice are
observed to form and dissipate throughout the lattice
[Fig.~\ref{ss:modelspatial}~(right)].  After sufficiently long times, we
observe the emergence of structured patterns having the symmetry of the
underlying lattice, with the behavior of agents belonging to a particular
structure being highly correlated.  Note that these patterns are dynamic,
being essentially concentric waves that emerge at the center and travel to
the  boundary of the region, which continually expands until it meets
another such pattern.  Where two patterns meet their progress is arrested
and their common boundary resembles a dislocation line.  In the asymptotic
limit, several such patterns fill up the entire system.  These patterns
indicate the growth of clusters with strictly correlated choice behavior.
The central site in these clusters act as the ``opinion leader'' for the
entire group.  This can be seen as analogous to the formation of ``cultural
groups'' with shared preferences~\cite{Axe97}.  It is of interest to note
that distributing $\lambda$ from a random distribution among the agents
disrupts the symmetry of the patterns, but we still observe patterns of
correlated choice behavior.  It is the global feedback ($\lambda \neq 0$)
which determines the formation of large connected regions of agents having
similar choice behavior. This is reflected in the order parameter, $\langle
| M | \rangle$, where $\langle \cdots \rangle$ indicates time averaging.
Fig.~\ref{ss:modeldistrn}~(right) shows the order parameter increasing with
$\lambda$ in both one and two dimensional lattices, signifying the
transition from a disordered state to an ordered state, where neighboring
agents have coordinated their choices.  

Our model seems to provide an explanation for the observed bimodality in a
large number of social or economic phenomena, e.g., in the distribution of
the gross income for movies released in theaters across the USA during the
period 1997-2003~\cite{Sinha04a}.  Bimodality in this context implies that
movies either achieve enormous success or are dismal box-office failures.
Based on the model presented here, we conclude that, in such a situation
the moviegoers' choice depends not only on their neighbors' choice, but
also on how well previous action based on such neighborhood information
agreed with media reports and reviews of movies indicating the overall or
community choice.  Hence, the case of $\lambda > 0$, indicating the
reliance of an individual agent on the aggregate information, imposes
correlation among agent choice across the community which leads to a
bimodal gross distribution. 

Based on a study of the rank distribution of movie earnings according to
their ratings~\cite{Vany02}, we further speculate that movies made for
children (rated G) have a significantly different popularity mechanism than
those made for older audiences (PG, PG-13 and R). The former show striking
similarity with the rank distribution curve obtained for $\lambda = 0$,
while the latter are closer to the curves corresponding to $\lambda > 0$.
This agrees with the intuitive notion that children are more likely to base
their choices about movies (or other products, such as toys) on the choice
of their friends or classmates, while adults are more likely to be swayed
by reports in mass media about the popular appeal of a movie.  This
suggests that one can tailor marketing strategies to different segments of
the population depending on the role that global feedback plays in their
decisions. Products whose target market has $\lambda = 0$ can be better
disseminated through distributing free samples in neighborhoods; while for
$\lambda > 0$, a mass media campaign blitz will be more effective. 

\section{Conclusions}
In this article we have primarily made an attempt to ascertain the general
empirical features inherent in many popularity phenomena. We observe that
the distribution of popularity in various contexts often exhibit long
tails, the nature of which seem to be either following a log-normal form or
a power law with the exponent $\alpha \simeq 1$ (Zipf's law). While the
log-normal distribution would arise naturally in any multiplicative
stochastic process, in the context of popularity it would be natural to
interpret it as a manifestation of the interplay of the multiple factors
involved in an agent making a decision to adopt a particular product or
idea. Further, there is no necessity for interactions among agents for this
particular distribution in popularity to be observed. On the other hand,
distributions with power law tails would seem to necessarily entail
inter-agent interactions, e.g., a process whereby agents follow the choice
of other agents, with a particular choice becoming more preferable if many
more agents opt for it~\footnote{In the economics literature, this is
referred to as positive externality~\cite{Arthur89}}. This is not
necessarily an irrational ``herding'' effect; for example, in the case of
popularity of cities, the larger the population of a city, the more likely
it is to attract migrants, owing to the larger variety of employment
opportunities.  Thus the very fact that more agents have chosen a
particular alternative may make that choice more preferable than others.
Seen in this light, the popularity distribution should show a log-normal
distribution in situations where individual quality preferences play an
important role in making a choice, while, in cases where the choice of
other agents is a paramount influence in the decision process of an agent,
Zipf's law should emerge~\footnote{Montroll \& Shlesinger~\cite{Montroll82}
have shown that a simple extension to multiplicative stochastic processes
can generate power-law tails from a log-normal distribution.  Recently,
Bhattacharyya et al~\cite{Chakrabarti05} have also proposed a very simple
model showing the asymptotic emergence of Zipf's law in the presence of
random interaction among agents; it is interesting in the context of our
statements here that, if the mean field theoretic arguments used in the
above paper are extended to the case of no interactions amongst agents,
they would suggest a log-normal distribution.}.  In either case, a
stochastic process is sufficient to generate the popularity distributions
seen in reality. This suggests that the emergence of popularity can be
explained entirely as an outcome of a sequence of chance events.

\acknowledgments{ We would like to thank S. Redner and M.~E.~J. Newman for
permission to use figures from their papers. SS would like to thank S.
Raghavendra for the many discussions during the early phase of the work on
popularity distributions. We would also like to thank D. Stauffer, B.~K.
Chakrabarti, M. Marsili and S. Bowles for helpful comments and suggestions
at various stages of this work. Part of this work was supported by the IMSc
Complex Systems Project funded by the DAE.}

\bibliography{chap1}
\end{document}